\documentclass[preprint,12pt]{elsarticle}

\usepackage[a4paper, left=1in, right=1in, top=1in, bottom=1in]{geometry}

\usepackage[cmex10]{amsmath}
\usepackage{amssymb}
\usepackage{cases}
\usepackage{graphicx}
\usepackage{mathtools}
\usepackage{multirow}
\usepackage{booktabs}
\usepackage{color}
\usepackage{algorithm}
\usepackage{makecell}
\usepackage{algpseudocode, setspace}

\usepackage{changepage}

\usepackage{url}
\usepackage{float}
\usepackage{verbatim}
\usepackage{soul}
\usepackage{amsthm,amsmath,amssymb}
\usepackage{mathrsfs}

\usepackage{bm}

\newcommand{\blue}[1]{\textcolor{blue}{#1}}

\soulregister\ref7 
\soulregister\cite7 

\DeclareMathOperator*{\argmax}{arg\,max}

\newtheorem{prop}{\textbf{Proposition}}
\usepackage[normalem]{ulem}

\allowdisplaybreaks

\journal{International Journal of Electrical Power and Energy Systems}

\begin{document}

        \begin{frontmatter}
                
                \title{Distribution Network Restoration with Mobile Resources Dispatch: A Simulation-Based Online Dynamic Programming Approach \tnoteref{Funding}}
                \tnotetext[1]{This work was jointly supported by National Key Research and Development Program of China (Grant Number 2022YFB2405500) and the Science and Technology Project of State Grid Corporation of China (Grant Number 52094023001H).}
                
                \author[1]{Mingxuan Li}
                
                \author[1]{Wei Wei\corref{cor1}}
                \ead{wei-wei04@mails.tsinghua.edu.cn}
                
                \author[2]{Yin Xu}
                
                \author[2]{Ying Wang}
                
                \author[3]{Shanshan Shi}
                
                \address[1]{State Key Laboratory of Power Systems, Department of Electrical Engineering, Tsinghua University, 100084, Beijing, China}
                \address[2]{School of Electrical Engineering, Beijing Jiaotong University, 100044, Beijing, China}
                \address[3]{State Grid Shanghai Municipal Electric Power Company, 200122, Shanghai, China}
                
                \cortext[cor1]{Corresponding author}
                
                \begin{abstract}
                        Dispatching mobile resources such as repair crews and mobile emergency generators is essential for the rapid restoration of distribution systems after extreme events. However, the restoration process is affected by various uncertain factors including repair time, road condition, and newly observed failures, necessitating online decision-making in response to real-time information. This paper proposes a simulation-based online dynamic programming approach to provide real-time restoration actions considering the dispatch of mobile resources. Using an index-based priority rule as the base policy, the remaining cumulative loss at the current state and a given action is evaluated from online simulation. As the base policy is explicit, the simulation is efficient. Then, the action leading to the minimum cumulative loss is chosen. It is proven that such a strategy improves the performance of base policy. The proposed policy adapts to real-time information updates and does not rely on offline training, so incurs no data and convergence-related issues, which is important in restoration tasks where the historical data of extreme events is rare. The rolling optimization approach may not meet the requirement of online use, because routing mobile resources gives rise to large-scale discrete optimization problems. Case studies on 123-bus and 8500-bus systems demonstrate that the proposed method achieves higher efficiency and better performance compared with rolling horizon optimization.   
                \end{abstract}
                
                
        \end{frontmatter}
        
        \section{Introduction}
        
        \subsection{Background}

        In recent years, extreme meteorological disasters have happened more frequently, posing unprecedented challenges to the security of distribution systems \cite{resilience_review_1}. Power outages in extreme weather conditions result in substantial economic losses, and system restoration requires coordination among various resources such as line switching, repair teams, backup units, and distributed energy resources. With advancements in technology and management, mobile emergency generators \cite{pre_disaster_SP_1} and mobile energy storage \cite{pre_disaster_RO_2} have been deployed as indispensable components in distribution network restoration \cite{MEG_hurricane_sandy}, attracting wide attention in recent research \cite{MES_6, pre_disaster_SP_3}. 

        \subsection{Literature Review}
        
        Existing work on power grid resilience considering mobile resources dispatch mainly focuses on modeling and decision support. The former category integrates vehicle routing with component repairing, spatial transportation of energy,  and grid restoration, yielding different discrete optimization problems. For example, a mixed-integer programming (MIP) model is proposed in \cite{MIP_MEG_1} for the scheduling of repair crews and mobile emergency generators, explicitly representing their travel and operational states. Schedules of electric buses are modeled in \cite{MIP_EB_1, MIP_EB_3}, considering both their public transit services and their participation in load supply. Traffic flow is modeled in \cite{MIP_road_2} to depict the detailed travel process of mobile resources. 
        These studies focus on detailed modeling of post-disaster restoration problems by incorporating various resources and factors, aiming to develop restoration strategies that are rigorously aligned with practical requirements. However, this also increases problem complexity and introduces diverse uncertainties, thereby posing significant challenges for decision-making algorithms.
        
     Post-disaster restoration involves multiple sources of uncertainty, such as unpredictable repair durations, road conditions, newly identified failures, and post-disaster load demands. Many of these uncertainties are inherently challenging to predict, with critical information often emerging dynamically during the restoration process. For instance, repairing a failure may reveal that the actual duration far exceeds initial estimates. In such situations, decisions need to be made promptly  in response to the evolving system state. Decision-making methods for post-disaster restoration under uncertainty can be broadly categorized into three types, as outlined below.
        
        The first type relies on two-stage optimization, such as stochastic programming and robust optimization. In this framework, the decision variables in the first stage are determined in advance, accounting for various possible realizations of future uncertainties. The second-stage variables are decided after observing the uncertainties. In \cite{Decomposed_SP_1}, a two-stage stochastic MIP is proposed to deal with the uncertainty of repair time and load demand when deciding repair schedules. Robust optimization is employed in \cite{RO_1} for the dispatch of repair crews and mobile power sources considering uncertainties in renewable output and load demand. In \cite{DRO_1} and \cite{DRO_2}, distributionally robust optimization is applied to handle uncertainties in renewable power output and outdoor temperature during restoration. In \cite{SP_PH_1}, a two-stage restoration strategy is proposed, where the first stage employs stochastic programming to handle uncertainties, and an improved progressive hedging algorithm is introduced to reduce the computational burden. A hybrid robust stochastic optimization model is proposed in \cite{RO_SP_1} to provide reliable decisions for microgrid operation. However, like all two-stage optimization methods, the above works implicitly assume that uncertainty information becomes available before the second-stage decisions are made, limiting online adaptability. When uncertainty unfolds period by period and decisions must be taken sequentially, the two-stage framework fails to satisfy the non-anticipativity constraint \cite{nonantcp}. As restoration problems scale up, the computational complexity of stochastic and robust optimization hinders real-time re-optimization, permitting only partial adjustments, such as real-time power flow. Consequently, in large-scale restoration problems with incrementally unfolding uncertainty, the two-stage framework struggles to accommodate online adjustments, potentially violating the non-anticipativity constraint and reducing its effectiveness in post-disaster restoration.

        The second type is rolling horizon optimization or model predictive control (MPC), where the decisions are extracted from a look-ahead problem based on newly updated forecasts. By employing time truncation and problem simplification, the optimization problem can be solved within a limited time, enabling rolling updates for restoration decisions. In \cite{MPC_SP_1}, MPC is applied to mobile power source scheduling considering uncertainty in road conditions and damaged lines. In \cite{MPC_2}, MPC is used to achieve dynamic restoration schedules utilizing distributed energy resources. Compared with two-stage methods, MPC includes prediction in the real-time stage and enjoys better adaptability. However, the performance of MPC heavily depends on prediction prediction accuracy. As noted earlier, historical data following extreme events is scarce compared to normal operating conditions. Moreover, predictions may span multiple sectors, including those beyond the power grid's control, complicating forecasting in network restoration. Furthermore, the schedule of mobile resources involves path planning entails discrete decisions, the time limit of deploying emergence actions may not be sufficient for solving large-scale MIP problems, which adds another dimension of difficulty to the application of MPC.
        
        The third type resorts to reinforcement learning (RL), where policies are trained offline and decisions are made online based on observed system states. By training value networks and policy networks, the RL agent can generate optimal decisions in real time. Conventional RL methods have been applied to post-disaster dispatch of energy storage and distributed generators, while recent studies have increasingly explored discrete decisions such as switch control. RL is applied in \cite{RL_4} and \cite{RL_topo_2} to achieve dynamic network reconfiguration. A curriculum-based RL method is proposed in \cite{RL_1} for the dispatch of generators and energy storage during restoration. In \cite{RL_2}, imitation learning is used to assist the training of the RL model for switch control. In \cite{RL_6} and \cite{RL_7}, RL models are trained to learn the optimal energization path of distribution networks.
        Reference \cite{RL_MESS} further applies RL to the scheduling of mobile energy storage.  
        In \cite{RL_RC}, RL is used  for the  scheduling of repair crews, where the action space is defined as different traveling directions, providing a way to model the routing of mobile resources. In \cite{DL_MCTS_RC}, neural networks are utilized to model the policy network for predicting action probabilities and the value network for estimating the value function, facilitating crew dispatch in distribution network restoration. The aforementioned studies demonstrate effective AI-based approaches for distribution network restoration. However, AI-based methods may encounter challenges when applied to large-scale restoration problems. High-dimensional state and action spaces make it challenging to train a robust agent that performs effectively in previously unseen fault scenarios. Moreover, the decision space for mobile resources might be dynamic; for instance, in different fault scenarios, repair teams may have varying repair targets. This variability complicates RL training, making it difficult for agents to explore and adapt to a changing decision space. In addition, the scarcity of data on extreme events could pose an additional concern.

        Table \ref{tab:review} provides a summary of the related research from various perspectives, where RES stands for renewable energy sources, RC represents repair crews, and MES refers to mobile energy sources. Regarding restoration resources, the table primarily highlights mobile resources due to the complexities of their scheduling during restoration. Common stationary resources, such as distributed power sources and switches, are not included.

        \begin{table}[h]
        \centering
        \scriptsize
        \renewcommand{\arraystretch}{1.2} 
        \caption{Summary of decision-making methods for post-disaster restoration under uncertainty}
        \label{tab:review}
        \begin{tabular}{ccccccc} 
                \toprule
                Method & Ref. & Uncertainty & \makecell{Mobile \\ Resource} &  \makecell{Require \\ Model \\ Training} & \makecell{Dependence on \\ Uncertainty \\ Prediction} \\
                \midrule
                \multirow{5}{*}{\makecell{Two-Stage \\ Optimization}} 
                & \cite{Decomposed_SP_1} &  Repair Time, Load Demand & RC & No & Medium \\
                &  \cite{RO_1} &  RES Output, Load Demand & RC, MES & No & Medium \\
                & \cite{DRO_1, DRO_2} &  RES Output, Load Demand & /  & No & Medium \\
                & \cite{SP_PH_1} &  RES Output, Load Demand & MES & No & Medium \\
                & \cite{RO_SP_1} &  \makecell{Repair/Travel Time, \\ RES Output, Load Demand} & RC & No & Medium\\
                \midrule
                \multirow{2}{*}{\makecell{Model \\ Predictive \\ Control}} 
                & \cite{MPC_SP_1} &  \makecell{Load Demand, Road Condition, \\ Fault Information} & MES & No & High \\
                & \cite{MPC_2} &  RES Output, Load Demand & / & No & High \\
                \midrule
                \multirow{4}{*}{\makecell{Reinforcement \\ Learning}} 
                & \cite{RL_4}-\cite{RL_1} &  RES Output & / & Yes & Low \\
                & \cite{RL_2} &  Fault Information & / & Yes & Low \\
                & \cite{RL_6} &  Switch Operability & / & Yes & Low \\
                & \cite{RL_MESS} &  \makecell{RES Output, Load Demand, \\ Travel Time} & MES & Yes & Low \\
                \midrule
                \multicolumn{2}{c}{Proposed Method} & \makecell{Repair Time, Fault Information, \\  RES Output, Load Demand} & RC, MES & No & Low \\ 
                \bottomrule
        \end{tabular}
        \end{table}

        \subsection{Research gap and contribution}
        
        In summary, existing approaches for decision-making in post-disaster restoration face several challenges, including insufficient data, unreliable predictions, computational inefficiency, and reliability concerns. This paper aims to develop an online decision method that makes full use of the mathematical model of the power grid, stochastic forecasts, and human experiences, and more importantly, derive a performance bound that can be theoretically proven.  Specifically, we propose a simulation-based online dynamic programming (ODP) approach for post-disaster restoration decision-making. While the ODP framework has been applied in other domains, such as robot scheduling for pipeline maintenance \cite{ODP_robot}, its implementation is inherently problem-specific. For the first time, we develop a customized simulation-based ODP algorithm tailored to the unique characteristics and constraints of power system restoration, ensuring its practical applicability and effectiveness in this context. The main contributions of this paper include
        
        (1) A tailored simulation-based ODP approach for post-disaster restoration. The proposed method accounts for multiple sources of uncertainty relevant to practical applications. An index-based priority rule serves as the base policy, with priorities for fault repair and target bus selection for mobile emergency generators calibrated based on experience. Given any system state, the remaining cumulative loss at the current state and a given action under the base policy is evaluated from online simulation. Among some promising actions, the one leading to the minimum cumulative loss is chosen. This method does not train value functions or policies offline. Instead, the state-action values are evaluated only for a small set of promising actions using the base policy. As the base policy is explicit, no optimization problem is solved, and the simulation is fast and can be conducted online. In this way, the curse-of-dimension issue is circumvented. The online action selection strategy resembles the Alpha-Zero program\cite{bertsekas_book} and is interpretable, as it adheres to Bellman's Principle of Optimality.

        (2) Proof of performance bound. It is proven that the ODP method is statistically no inferior compared to the base policy, which follows guidelines or rules that are determined according to human experiences. In contrast, conventional RL methods do not have such a guarantee. Case study results indicate that ODP significantly reduces the optimality gap compared to directly applying the base policy.
        
        The remainder of the paper is organized as follows. Section \ref{section: prob_form} formulates the online decision problem for post-disaster distribution network restoration. Section  \ref{section: solution} introduces the proposed simulated-based online dynamic programming method. Section \ref{section: case study} conducts case studies on IEEE 123-bus and 8500-bus systems. Section \ref{section: conclusion} concludes the paper.
        
        \section{Problem Formulation} \label{section: prob_form}
        Distribution networks are often operated under radial topology. However, distribution lines could be damaged in extreme weather, disrupting the connectivity of the network. Due to imbalances between generation and load, certain network islands may experience power shortages. After the disaster, the distribution network needs to restore loads through network reconfiguration and dispatch of mobile resources. The overall restoration procedure is quite complicated. To ease the model setup, we made the following assumptions and simplifications without losing the main characteristics of the restoration problem:
        
        {\bf 1) Damaged Components}. In a distribution network, substations are well protected while distribution lines are vulnerable components that could be destroyed by hurricanes. So we mainly consider distribution line outage in this work, which is a common focus in restoration research \cite{MIP_line_repair_1, MIP_line_repair_2}. Other damaged components can be modeled similarly.

        {\bf 2) Uncertain factors}. Various uncertainties can affect the restoration process, among which we select the most important ones in the post-disaster stage: repair time and undiscovered faults. The precise repair duration for a damaged line remains unknown until a repair crew visits and inspects it \cite{uncertain_repair_time_2}; the prior distribution of uncertain repair time can be learned from experience or historical data. In practice, we may not be able to discover all faults once after the disaster ends, and some new faults may be reported by users or patrol drones as time goes by. While the former has been addressed in prior research, the latter is rarely incorporated due to the absence of a concrete mathematical model.
        
        {\bf 3) Mobile resources}. We mainly consider mobile emergency generators, but mobile storage can be included as well. Once a mobile emergency generator or a repair team begins its journey or repair work, the process cannot be interrupted until completion. However, their future routes can be updated online according to real-time information even if they are currently occupied, which is a key advantage of the proposed online algorithm. Additionally, mobile emergency generators carry sufficient fuel, ensuring an adequate supply throughout the restoration period.

        Based on the above assumptions, a post-disaster restoration model can be established. This section first introduces the deterministic MIP model commonly used in existing research and analyzes its limitations in uncertain post-disaster scenarios.  Building on these considerations, an online decision-making model for emergency restoration is then proposed to support dynamic decision adjustments.

        \subsection{Deterministic Restoration Model} \label{section:MIP}

        The deterministic restoration model employs MIP to model the participation of mobile resources in distribution network restoration. The following sections present dispatch models for repair crews, mobile emergency generators, and the distribution network, leading to the formulation of a comprehensive MIP model for emergency restoration.

        \subsubsection{Operation of the Distribution Network} 
        
        The power flow constraints of the distribution network can be modeled by the linearized DistFlow model \cite{DistFlow}, as follows.
        \begin{subequations} \label{eq:DN_power}
                \begin{gather}
                z_{\ell, t} \le \sum_{\tau=1}^{t} h_{\ell, \tau} , \forall \ell \in \Omega^{FL}   \label{eq:DN_line_1}\\
                z_{\ell, t} = 1 , \forall \ell \notin \{\Omega^{FL} \cup \Omega^{SW}\} , \forall t   \label{eq:DN_line_2} \\
                p_{i, t}^{G} + \sum_{\mathclap{\forall (k, i) \in \Omega^L}} p_{ki, t}^{\ell} = p_{i, t}^{D}+\sum_{\mathclap{\forall (i, j) \in \Omega^L}} p_{ij, t}^{\ell}, \forall i, \forall  t \label{eq:DN_3}\\
                q_{i, t}^{G} + \sum_{\mathclap{\forall (k, i) \in \Omega^L}} q_{ki, t}^{\ell} = q_{i, t}^{D}+\sum_{\mathclap{\forall (i, j) \in \Omega^L}} q_{ij, t}^{\ell}, \forall i, \forall t \label{eq:DN_4}\\
                p_{i, t}^{G} = p_{i, t}^{DG}+p_{i, t}^{R}+p_{i, t}^{M} + p_{i, t}^{ES,dis} -  p_{i, t}^{ES,ch} \label{eq:DN_pg} \\
                q_{i, t}^{G} = q_{i, t}^{DG}+q_{i, t}^{R}+q_{i, t}^{M} + q_{i, t}^{ES,dis} -  q_{i, t}^{ES,ch} \label{eq:DN_qg} \\
                \begin{aligned}
                &\left| V_{j, t}-V_{i, t} + ({R_{\ell} p_{ij, t}^{\ell}+X_{\ell} q_{ij, t}^{\ell}})/{V_0} \right| \\
                & \qquad \qquad \qquad  \leq  M \left(1 -z_{\ell, t}\right), \forall \ell, \forall t  
                \end{aligned} \label{eq:DN_5} \\
                V_i^l \leq V_{i, t} \leq V_i^u, \forall i, t \label{eq:DN_7}\\
                0 \leq p_{i, t}^{D} \leq P_{i,t}^D, \  0 \leq q_{i, t}^{D} \leq Q_{i,t}^D, \forall i, \forall t \label{eq:DN_load} \\
                0 \leq p_{i, t}^{DG} \leq \overline{{P}_{i}^{DG}}, \  0 \leq q_{i, t}^{DG} \leq \overline{{Q}_{i}^{DG}}, \forall i, \forall t \label{eq:DN_8}\\
                0 \leq p_{i, t}^{R} \leq \overline{{P}_{i,t}^{R}}, \  0 \leq q_{i, t}^{R} \leq \overline{{Q}_{i,t}^{R}}, \forall i, \forall t \label{eq:DN_9}\\
                E^{ES}_{i,t+1} = E^{ES}_{i,t} - \frac{1}{\eta_i}P^{ES,dis}_{i,t}+\eta_i P_{i,t}^{ES,ch} , \forall i,t \label{eq:DN_es_e1}\\
                E_{i, min}^{ES} \le  E_{i,t}^{ES} \le E_{i, max}^{ES}, \forall i, \forall t \label{eq:DN_es_e2} \\
                0 \le P^{ES, dis}_{i,t} \le \overline{P^{ES}_{i}}, 0 \le P^{ES, ch}_{i,t} \le \overline{P^{ES}_{i}}, \forall i, \forall t \label{eq:DN_es_p} \\
                -z_{\ell, t} \overline{P_{\ell}^L}  \leq p_{\ell, t}^{L} \leq z_{\ell, t} \overline{P_{\ell}^L}, \forall \ell, \forall t \label{eq:DN_10}\\
                -z_{\ell, t} \overline{Q_{\ell}^L} \leq q_{\ell, t}^{L} \leq z_{\ell, t} \overline{Q_{\ell}^L}, \forall \ell, \forall  t \label{eq:DN_11}
                \end{gather}
        \end{subequations}
        where $\Omega^{SW}$ is the set of switchable lines; $p_{i,t}^{G}$ represents the total active power generation at bus $i$ at time $t$, including the output from distributed generation $p_{i,t}^{DG}$, renewable energy $p_{i,t}^{R}$, mobile generators $p_{i,t}^{M}$, and energy storage discharging/charging power $p_{i,t}^{ES,dis} /  p_{i,t}^{ES,ch}$. Similarly, the total reactive power generation is denoted by $q_{i,t}^{G}$.
        ; $R_\ell/X_\ell$ is the resistance$/$reactance of line $\ell$; $M$ is a large constant; $P_{i,t}^D$ and $Q_{i,t}^D$ are the active and reactive demands at bus $i$ in period $t$; $\overline{{P}_{i}^{DG}}$ and $\overline{{Q}_{i}^{DG}}$ are the maximum active and reactive power generation at bus $i$; $\overline{{P}_{i,t}^{R}}$ and $\overline{{Q}_{i,t}^{R}}$ are the maximum active and reactive power output of renewable resource at bus $i$ in period $t$, depending on real-time weather condition; $E_{i,t}^{ES}$ denotes the energy stored in storage $i$ at time $t$, with upper and lower limits of $E_{i,max}^{ES}$ and $E_{i,min}^{ES}$. $\overline{P_i^{ES}}$ is the maximum charging and discharging power of storage $i$, and $\eta_i$ represents its efficiency; $\overline{P_{\ell}^{L}}$ and $\overline{Q_{\ell}^{L}}$ are maximum active and reactive power flow in line $\ell$. Constraints (\ref{eq:DN_line_1})-(\ref{eq:DN_line_2}) indicate the working status of faulty lines and normal non-switchable lines, respectively; (\ref{eq:DN_3})-(\ref{eq:DN_4}) are power balance equations; constraints (\ref{eq:DN_pg})-(\ref{eq:DN_qg}) detail the components of power generation at each bus; constraint (\ref{eq:DN_load}) prescribes that the supplied load cannot exceed the demand; constraint (\ref{eq:DN_5}) models the voltage drop along a distribution line; constraint (\ref{eq:DN_7}) defines the feasible interval of voltage magnitudes; Constraints (\ref{eq:DN_8})-(\ref{eq:DN_9}) limit the output of distributed generators and renewable energy; constraints (\ref{eq:DN_es_e1})-(\ref{eq:DN_es_p}) are energy storage operation constraints ; constraints (\ref{eq:DN_10})-(\ref{eq:DN_11}) limit the power flow of each line considering their working status.
        
        The radial topology can be modeled by single-commodity flow constraints \cite{radial_SCF}, assuming that there are virtual root nodes in the graph with fictitious resources, serving the other nodes with unit fictitious demand, yielding:
        \begin{subequations} \label{eq:DN_radial}
                \begin{gather}
                \sum_{{\forall (k,i)\in\Omega^L}} f_{ki,t}^\ell - \sum_{{\forall (i,j)\in\Omega^L}} f_{ij,t}^\ell = 1, \ \forall i \in \Omega^B \backslash\mathcal{V}, \label{eq:rad_1} \\
                -N_{B}z_{\ell,t} \le f_{ij,t}^\ell \leq N_B z_{\ell,t}, \forall \ell \in \Omega^L  \label{eq:rad_2}\\
                \sum\nolimits_{\ell\in \Omega^L}  u_\ell=N_B-N_{\mathcal{V}} \label{eq:rad_3}
                \end{gather}
        \end{subequations}
        where $f_{ij,t}^\ell$ represents the fictitious flow of line $\ell$ in period $t$;  $\mathcal{V}$ is the set of virtual root nodes. Constraint (\ref{eq:rad_1}) models the flow balance for each non-root node; constraint (\ref{eq:rad_2}) indicates there is no flow if the line is disconnected; constraint (\ref{eq:rad_3}) restricts the number of working lines in a radial network.
        
        \subsubsection{Dispatching Repair Crews}
        
        Multiple repair crews depart from their depots and proceed to each fault location for repairing work until all faults are fixed. The constraints for repair crew routing are as follows \cite{Decomposed_SP_1}:
        \begin{subequations} \label{eq:RC_route}     
                \begin{gather}  
                \sum_{\forall m \in \Omega^{FL}} x^R_{o_{c}, m, c} = 1,
                \sum_{\forall m \in \Omega^{FL}} x^R_{m, e_{c}, c} = 1, \forall c \label{eq:RC_route_2}\\
                \sum_{\forall n \in \Omega^{FL}} x^R_{m, n, c}-\sum_{\forall n \in \Omega^{FL}} x^R_{n, m, c} = 0, \forall c,\forall m \in \Omega^{FL} \label{eq:RC_route_3}\\
                \begin{split}
                | \zeta_{n, c} - \zeta_{m, c} - \mathcal{R}_m - \mathcal{T}^R_{m, n} | \le \left(1-x^R_{m, n, c}\right)  M, \\
                \forall m \in \Omega^{FL}, \forall n \in \Omega^{FL}\backslash m, \forall c \label{eq:RC_arrive_2}
                \end{split}  \\
                \sum_{\forall t}  h_{m, t} = 1, \forall m \in \Omega^{FL} \label{eq:RC_recovery_1}\\
                \sum_{\forall t} (t\cdot h_{m, t}) \geq \sum_{\forall c}(\zeta_{m, c}+\mathcal{R}_m \sum_{\mathclap{\forall n \in \Omega^{FL}}} x^R_{m, n, c}), \ \forall m \in \Omega^{FL} \label{eq:RC_recovery_2} \\
                \begin{split}
                \sum_{\forall t} (t\cdot h_{m, t}) \leq \sum_{\forall c}(\zeta_{m, c}+\mathcal{R}_m \sum_{\forall n \in \Omega^{FL}} x^R_{m, n, c}) \\ +1-\epsilon, \forall m \in \Omega^{FL} \label{eq:RC_recovery_3}  
                \end{split}        
                \end{gather}      
        \end{subequations}
        where binary variable $x^R_{m, n, c}=1/0$ indicates whether crew $c$ travels from the origin line $m$ to the destination line $n$ in the route or not; $\Omega^{FL}$ is the set of faulty lines; $\zeta_{n, c}$ is the time when repair crew $c$ arrives at line $n$; $\mathcal{R}_{m}$ is the repair time of line $m$; $\mathcal{T}^R_{m, n}$ is the travel time; $M$ is a large constant; Binary variable $h_{m, t}=1/0$ indicates whether line $m$ is restored in period $t$ or not.  Constraint (\ref{eq:RC_route_2}) shows that each repair crew $c$ departs from $o_c$ and eventually returns to $e_c$, where $e_c=o_c$ if repair crews are required to return to the depots; constraint (\ref{eq:RC_route_3}) stipulates path continuity; constraint (\ref{eq:RC_arrive_2}) calculates the arrival time at line $n$; constraint (\ref{eq:RC_recovery_1}) shows that each line is restored at a specific time; constraints (\ref{eq:RC_recovery_2})-(\ref{eq:RC_recovery_3}) specify the restoration time of each line. 
        
        \subsubsection{Dispatching Mobile Emergency Generators}
        After a disaster, mobile emergency generators will travel to access points in the distribution network for power supply. Constraints of mobile emergency generator scheduling include \cite{MIP_MEG_1}:
        \begin{subequations} \label{eq:MEG_route}
                \begin{gather}
                \sum_i\sum_j x_{g,t,i,j}^M = 1, \forall t,\forall g, \forall i,\forall j \label{eq:MEG_route_1}\\
                x_{g,t,i,i}^M + x_{g,t-1,j,j}^M \le 1, \forall t,\forall g, \forall i,\forall j \label{eq:MEG_route_2}\\
                x_{g,t,i,j}^M - x_{g,t-1,i,j}^M  \le x_{g,t-1,i,i}^M, \forall t,\forall g, \forall i,\forall j \label{eq:MEG_route_3}\\
                x_{g,t,j,j}^M - x_{g,t-1,j,j}^M  \le \sum\nolimits_i x_{g,t-1,i,j}^M, \forall t,\forall g, \forall j \label{eq:MEG_route_4}\\
                \sum_{\tau=t}^{\mathclap{t+\mathcal{T}_{i j}^M-1}} x_{g, \tau, i, j}^M \geq \mathcal{T}_{i j}^M (x_{g, t, i, j}^M-x_{g, t-1, i, j}^M), \forall t,\forall g, \forall i,\forall j \label{eq:MEG_route_5} 
                \\
                x_{g, t+\mathcal{T}_{i j}^M, j, j}^M \geq x_{g, t, i, j}^M-x_{g, t-1, i, j}^M, \forall t,\forall g, \forall i,\forall j \label{eq:MEG_route_6} 
                \end{gather}
        \end{subequations}
        where $x_{g,t,i,j}^M=1$ indicates generator $g$ is traveling from bus $i$ to bus $j$ in period $t$ if $i\neq j$, or parking at bus $i$ if $i=j$. $\mathcal{T}_{ij}^M$ denotes the travel time between buses $i$ and  $j$; constraint (\ref{eq:MEG_route_1}) indicates that each generator is either traveling or supplying power in each period; Constraint (\ref{eq:MEG_route_2}) ensures that generators cannot appear at a bus without traveling; Constraint (\ref{eq:MEG_route_3}) states that a generator can depart from bus $i$  and travel to bus $j$ only if it is currently located at bus $i$; Constraint (\ref{eq:MEG_route_4}) shows that a generator may arrive at bus $j$ only after departing from another bus $i$; Constraints (\ref{eq:MEG_route_5})-(\ref{eq:MEG_route_6}) model the traveling state and the arrival time based on travel time $\mathcal{T}_{ij}^M$.
        
        Based on the parking state, the power supply of the mobile emergency generator can be modeled as 
        \begin{subequations} \label{eq:MEG_power}
                \begin{gather}
                0 \le p^M_{i,t} \le \sum_{g\in \Omega^M} \overline{P_g^{M}} x^M_{g,t,i,i}, \forall i\in \Omega^{AP}, \forall t \\
                0 \le q^M_{i,t} \le \sum_{g\in \Omega^M} \overline{Q_g^{M}} x^M_{g,t,i,i}, \forall i\in \Omega^{AP}, \forall t
                \end{gather}
        \end{subequations}
        where $P_g^{M, ub}$ and $Q_g^{M, ub}$ are the maximum active and reactive power output of generator $g$. Constraint (\ref{eq:MEG_power}) indicates that a generator can supply power when it is parking at some bus.

        \subsubsection{MIP Formulation of Restoration}
        Combining all the above constraints, the restoration problem is cast as
        \begin{equation} \label{eq:opt_restore_MIP}
        \begin{split}
        \min \sum_{t=1}^{T} \sum_{i=1}^{N_B} c_{i} (P_{i}^D-p_{i,t}^D) \Delta t \\
        s.t. (\ref{eq:DN_power}), (\ref{eq:DN_radial}) , (\ref{eq:RC_route}), (\ref{eq:MEG_route}), (\ref{eq:MEG_power}), 
        \end{split}
        \end{equation}
        
        The MIP problem (\ref{eq:opt_restore_MIP}) can be solved using off-the-shelf solvers, providing restoration strategies for all time periods. However, the large number of integer variables significantly reduces solving efficiency for large-scale problems, and the model lacks adaptability to uncertainties or parameter updates. To address this limitation, we reformulate the problem as an online decision model that enables real-time adjustments, detailed in the following sections.

        \subsection{Online Decision Model} \label{section:Model_OL}
        
        In the online decision model for the restoration problem, system states and restoration decisions can be updated at each time step. Restoration decisions are made in response to observable information without precise forecasts. 
        
        \subsubsection{Description of Distribution Network}
        
        Consider a distribution network with buses in set $\Omega^B$ and lines in set $\Omega^L$; $p^{G}_{i,t}/q^{G}_{i,t}$ is active$/$reactive power generation at bus $i \in \Omega^B$ in period $t$, $p_{i, t}^{D}/q_{i, t}^{D}$ represents active$/$reactive load that is supplied in period $t$; $V_{i,t}$ is squared bus voltage magnitude; $p_{ij,t}^\ell/ q_{ij,t}^\ell$ is active$/$reactive power flow in line $\ell \in \Omega^L$ connecting bus $i$ and bus $j$. Binary variable $z_{\ell, t}$ indicates the open$/$closed status of line $\ell$. These variables satisfy power flow constraints in Section \ref{section:MIP}.
        
        To establish an online model, variables of the distribution network are classified into three groups: system state
        \begin{equation*}
        s_t^{DN} = \{\Omega^{FL}_t, r_{\ell,t} \}, \forall \ell\in \Omega^{FL}_t
        \end{equation*}
        is observed at the beginning of period $t$, where $\Omega^{FL}_t$ represents the set of damaged lines observed before period $t$;  $r_{\ell,t}$ denotes the cumulative time spent on repairing line $\ell$ in period $t$, which is used to determine whether line $\ell$ has been restored. After observing the state in period $t$, the operator attempts to restore loads through action
        \begin{equation*}
        a_{t}^{DN} = \{p^G_{i,t}, q^G_{i,t}, p^D_{i,t}, q^D_{i,t}, z_{\ell,t}\}, \ \forall i\in{\Omega^B} , \forall \ell \in \Omega^L
        \end{equation*}
        where active power generation  $p^G_{i,t}=[p^{DG}_{i,t},  p^R_{i,t}, p^{ES}_{i,t}, p_{i,t}^M]$ includes  outputs from distributed generator $p^{DG}_{i,t}$, renewable plant $p^{R}_{i,t}$, energy storage $p^{ES}_{i,t}$, and mobile emergency generator $p_{i,t}^M$, so does reactive power output $q^G_{i,t}=[q^{DG}_{i,t}, q^{R}_{i,t}, q^{ES}_{i,t}, q_{i,t}^M]$; Supplied demand  $p^D_{i,t}$ and $q^D_{i,t}$ are determined in $a_t^{DN}$, which directly affects loss of demand. Other variables are included in the algebraic variable \begin{equation*}
        y_t^{DN} = \{V_{i,t},p_{ij,t}^\ell, q_{ij,t}^\ell\},~
        \forall i\in{\Omega^B} , \forall \ell \in \Omega^L
        \end{equation*} 
        
        The action space $\mathcal A^{DN}_t$ of $a_t^{DN}$ is affected by the locations of mobile emergency generators through (\ref{eq:MEG_power}), power flow constraints in (\ref{eq:DN_power}), where the algebraic variable $y_t^{DN}$ is internally determined by power flow and topology constraint (\ref{eq:DN_radial}). Therefore, the action set is defined as
        \begin{equation*}
        \mathcal A_{t}^{DN} = \left\{a_t^{DN} ~\middle|~ \begin{aligned}
        & \exists y^{DN}_t: (a^{DN}_t,s^{DN}_t,y^{DN}_t) \\
        &\mbox{satisfying } (\ref{eq:MEG_power}), (\ref{eq:DN_power}), (\ref{eq:DN_radial})
        \end{aligned} \right\}
        \end{equation*}

        \subsubsection{Description of Repair Crew}

        Routing repair crews involves assigning faulty lines to each crew and determining the repair order. Rather than determining the entire sequence at once, the online policy selects the next target for team $c \in \Omega^R$ in each period. The state $s_t^{RC}$ and action $a_t^{RC}$ for repair crews are defined as follows:
        \begin{equation*}
        s_t^{RC} = \{\alpha_{c,t}^{R}, u_{c,t}^{R},  \tau_{c,t}^{R}\}, \forall c \in \Omega^{R}
        \end{equation*}
        \begin{equation*}
        a_{t}^{RC} = \{\beta_{c,t}^{R}  \}, \ \forall c\in{\Omega^R}
        \end{equation*}
        where $\alpha_{c,t}^{R}$ is the line that crew $c$ is repairing in period $t$, which is $\emptyset$ if crew $c$ is not actively repairing a line; $u_{c,t}^{R} =1/0$ means that crew $c$ is repairing a fault$/$traveling; $\tau_{c,t}^{R}$ is the remaining time for repair crew $c$ traveling to the next target; $\beta_{c,t}^{R}$ is the target line for repair crew $c$. If crew $c$ is actively repairing a line, $\beta_{c,t}^{R}$ represents the next line it plans to visit. 
        
        As each repair crew is assigned different lines to avoid conflicting targets, the action space for $a_t^{RC}$ is 
        \begin{equation*}
        \mathcal A_{t}^{RC} = \left\{ a_{t}^{RC} ~\middle|~ 
        \begin{aligned}
        & \beta_{c,t}^R \in\Omega^{FL}_t , \ \forall c\in \Omega^R \\
        & \beta_{c_1,t}^R \ne \beta_{c_2,t}^R , \forall c_1, c_2\in \Omega^R
        \end{aligned}  \right\}
        \end{equation*}

        \subsubsection{Description of Mobile Emergency Generator}
        
        Mobile generators can travel to different access points and supply isolated islands. The state $s_t^{MEG}$ and action $a_t^{MEG}$ of a mobile emergency generator $g \in \Omega^{M}$ are defined as: 
        \begin{equation*}
        s_t^{MEG} = \{\alpha_{g,t}^{M}, u_{g,t}^{M}, \tau_{g,t}^{M}\}, \forall g \in \Omega^{M}
        \end{equation*}
        \begin{equation*}
        a_{t}^{MEG} = \{ \beta_{g,t}^{M} \} \ \forall g\in {\Omega^M} 
        \end{equation*}
        where $\alpha_{g,t}^{M}$ denotes the bus to which mobile emergency generator $g$ is currently connected; it is $\emptyset$ if the generator is in transit; $u_{g,t}^{M} = 1/0$ means that mobile emergency generator $g$ is generating power$/$traveling in period $t$; $\tau_{g,t}^{M}$ represents the remaining travel time to reach the next target bus $\beta_{g,t}^{M}$. The action space for $a_t^{MEG}$ is
        \begin{equation*}
        A_t^{MEG} = \left\{a_t^{MEG} \mid \beta_{g,t}^{M}\in \Omega^{AP}, \forall g \in \Omega^M \right\}
        \end{equation*}
        where $\Omega^{AP}$ is the set of accessible points. The dispatched power $(p_{i,t}^M, q_{i,t}^M)$ is included in $a_t^{DN}$ 
        to ensure coordination with other generation resources, maintaining compliance with the distribution network's power flow constraints.

        \subsubsection{State Transition}
        
        Among the aforementioned actions, $a_t^{DN}$ influences the current load loss using the power flow model, while $a_t^{RC}$ and $a_t^{MEG}$ affect the future states. States and actions of the restoration problem collect those defined above:  
        \begin{equation*}
        \begin{gathered}
        s_t=\{s_t^{DN}, s_t^{RC}, s_t^{MEG}\} \\
        a_t=\{a_t^{DN}, a_t^{RC}, a_t^{MEG}\}
        \end{gathered}
        \end{equation*}
        The observed uncertainty at the beginning of period $t$ is
        \begin{equation*}
        \xi_{t}=\{\hat{\mathcal{R}}_{\ell,t}, \Omega^{NL}_{t} , \overline{P_{i,t}^{R}}, {P_{i,t}^{D}}\}, \ \forall \ell \in \Omega^{FL}_t, \forall i \in{\Omega^B}
        \end{equation*}
        where $\hat{\mathcal{R}}_{\ell,t}$ is the repair time for line $\ell$  estimated in period $t$. If a repair crew has reached line $\ell$, the exact repair time becomes clear based on a detailed assessment;  otherwise, $\hat{\mathcal{R}}_{\ell,t}$ remains uncertain in period $t$. $\Omega^{NL}_t$ denotes newly discovered faults in period $t$. $\overline{{P}_{i,t}^{R}}$ is the maximum output of renewable plants at bus $i$ in period $t$, while ${P_{i,t}^{D}}$ is the active power demand of bus $i$ in period $t$.

        Based on the above notations, the transition function $s_{t+1} =  f(s_t, a_t, \xi_{t})$ includes the following equations:
        \begin{subequations}{\label{State_Trans}}
                \begin{align}
                r_{\ell,t+1} & = r_{\ell, t} + \sum_{c\in \Omega^R}  u_{c,t}^{R}\mathbb{I}(\alpha_{c,t}^{R}=\ell), \forall \ell\in \Omega^{FL}_t \label{Tr_r} \\
                {\Omega}^{FL}_{t+1} & = \Omega^{FL}_{t} \cup {\Omega^{NL}_{t}}\ /\ \Omega^{RL}_{t+1} \label{Tr_FL} \\
                \tau_{c, t+1}^{R} & = \begin{cases} 
                max\{0, \tau_{c, t}^{R} - 1 \}, \  \mbox{if} \  u_{c,t}^{R}=0 \\
                \mathcal{T}^R(\alpha_{c,t}^{R}, \beta_{c,t}^{R}), \quad\ \ \mbox{if} \  u_{c,t}^{R}=1 
                \end{cases} , \forall c
                \label{Tr_tau_RC}
                \\
                \alpha_{c,t+1}^{R} & = \begin{cases}
                \emptyset, \quad \ \mbox{if} \  u_{c,t}^{R}=0  \land  \tau_{c, t+1}^{R} > 0 \\
                \beta_{c,t}^{R}, \ \mbox{if} \  u_{c,t}^{R}=0  \land  \tau_{c, t+1}^{R} \le 0 \\
                \alpha_{c,t}, \ \mbox{if} \  u_{c,t}^{R}=1 \ 
                \end{cases}, \forall c
                \label{Tr_n_RC}
                \\
                \begin{split}
                u_{c,t+1}^{R} & = \mathbb{I}
                \begin{pmatrix}
                \left(\alpha_{c,t+1}^{R} = \beta_{c,t}^{R} \lor u_{c,t}=1\right)  \\   \ \land \left(r_{(\alpha_{c,t+1}^{R}), t+1} < \hat{\mathcal{R}}_{(\alpha_{c,t+1}^{R}),t}\right) \end{pmatrix}
                ,\ \forall c \label{Tr_u_RC}
                \end{split}  
                \\
                \tau_{g, t+1}^{M} & = \begin{cases}
                max\{0, \tau_{g, t}^{M} - 1 \}, \ \mbox{if} \  u_{g,t}^{M}=0 \\
                \mathcal{T}^M(\alpha_{g,t}^{M}, \beta_{g,t}^{M}), \quad\ \mbox{if} \  u_{g,t}^{M}=1
                \end{cases} , \forall g
                \label{Tr_tau_MEG}
                \\
                \alpha_{g,t+1}^{M} & = \begin{cases}
                \emptyset, \quad\ \mbox{if} \  u_{g,t}^{M}=0 \land  \tau_{g, t+1}^{M} > 0 \\
                \beta_{g,t}^{M}, \ \mbox{if} \  u_{g,t}^{M}=0 \land  \tau_{g, t+1}^{M} \le 0 \\
                \alpha_{g,t}^{M}, \ \mbox{if} \  u_{g,t}^{M}=1
                \end{cases}, \forall g
                \label{Tr_n_MEG} \\
                u_{g,t+1}^{M} & = \mathbb{I}(\alpha_{g,t+1}^{M} = \beta_{g,t}^{M}), \forall g \label{Tr_u_MEG}  
                \end{align}   
        \end{subequations}
        where $\Omega^{RL}_{t+1}  = \{\ell|\  r_{\ell,t+1}\ge \mathcal{R}_{\ell,t+1} \}$ is the set of lines that have been restored by period $t+1$; $\mathcal{T}^R(\alpha_{c,t}^{R}, \beta_{c,t}^{R})$ is the time that repair crew $c$ travels from the current line $ \alpha_{c,t}^{R}$ to the target line  $\beta_{c,t}^{R}$; $\mathcal{T}^M(\alpha_{g,t}^{M}, \beta_{g,t}^{M})$ is the time that mobile emergency generator $g$ travels from the current bus $\alpha_{g,t}^{M}$ to the target bus $\beta_{g,t}^{M}$. Indicator function $\mathbb{I}(x)=1/0$ if $x$ is true$/$false. 
        
        Equation (\ref{Tr_r}) updates the cumulative repairing time of each line. Equation (\ref{Tr_FL}) shows that the set of faulty lines is updated by adding newly discovered lines and removing repaired ones.  Equations  (\ref{Tr_tau_RC})-(\ref{Tr_tau_MEG}) calculate the remaining travel time of repair crews and mobile emergency generators to reach their targets, respectively. Equations (\ref{Tr_n_RC})-(\ref{Tr_n_MEG}) update the current location of repair crews and mobile emergency generators. Equation (\ref{Tr_u_RC}) updates the travelling$/$repairing status of each crew, indicating that a crew is conducting repair work if it has just arrived at the target or if the current repair work has not been finished yet, and is traveling otherwise. Equation (\ref{Tr_u_MEG})  shows whether the mobile emergency generator $g$ is staying at the target bus or traveling.

        \subsubsection{Optimal Policy for Restoration}
        
        The unsatisfied demand in period $t$ is 
        \begin{equation*}
        C(s_t,a_t) = \sum_{i=1}^{N_B} c_{i} (P_{i,t}^D-p_{i,t}^D) \Delta t
        \end{equation*}
        where $c_i$ is the cost of load shedding at bus $i$; $P_{i,t}^D$ is the active load demand at bus $i$ in period $t$, while $p_{i,t}^D$ is the supplied demand; and $\Delta t$ is the duration of period $t$. Restoration aims to minimize the expected cumulative load curtailment, yielding
        \begin{equation}
        \label{eq:Restoration-multiperiod}
        \begin{aligned}
        \min_{\{a_t(\cdot)\}_{t=1}^T}~~ & \mathbb E \left[\sum_{t=1}^T  C(s_t,a_t(\cdot)) \right] \\
        \mbox{s.t.} \qquad & s_{t+1} =  f(s_t, a_t, \xi_{t})  
        \end{aligned}
        \end{equation}
        where policy $a_t(s_t,\xi_{[t]})$ is a function of system state $s_t$ and observation history $\xi_{[t]}=\{\xi_1,\cdots,\xi_t \}$; the expectation is taken over probability distribution of $\xi$. Problem (\ref{eq:Restoration-multiperiod}) entails optimization over functions, it cannot be solved exactly.   
        
        Following the paradigm of dynamic programming, define the state-action value function (also called Q-function) 
        \begin{equation}
        \label{eq:Q-fun}
        Q(s_t, a_t) =C(s_t, a_t) + \min _{a_{t+1:T}}\mathbb{E} \left[ \sum_{\tau=t+1}^T C(s_\tau, a_\tau(\cdot))\right]
        \end{equation}
        where $a_t$ is a given value, then the action in period $t$ is 
        \begin{equation} \label{eq:Q_action}
        a_t = \arg \min_{a_t \in \mathcal A_t} Q(s_t,a_t) 
        \end{equation}
        
        In reinforcement learning, the Q-function $Q(s_t, a_t)$ is typically trained using neural networks, with the expectation in (\ref{eq:Q-fun}) being approximated using large amounts of data. However, training such an agent might be challenging for large-scale restoration problems. To address this, we propose an alternative online decision-making approach in the next section that does not rely on pre-trained models. The proposed method leverages online simulation to evaluate candidate actions in real time, where $Q(s_t, a_t)$ is  efficiently evaluated at the current state $s_t$ and  some promising actions.  Most importantly, evaluating $Q(s_t, a_t)$ online can include uncertainty that is difficult to model offline, making the policy more adaptive to the changing environment. 
        
        \section{Online dynamic programming} \label{section: solution}
        
        This section begins by introducing the overall framework, followed by the technical details of the method.
        
        \subsection{Illustration of Simulation-based ODP} \label{subsection: ODP}

        \begin{figure}[!h]
                \centering
                \includegraphics[width=0.6\linewidth]{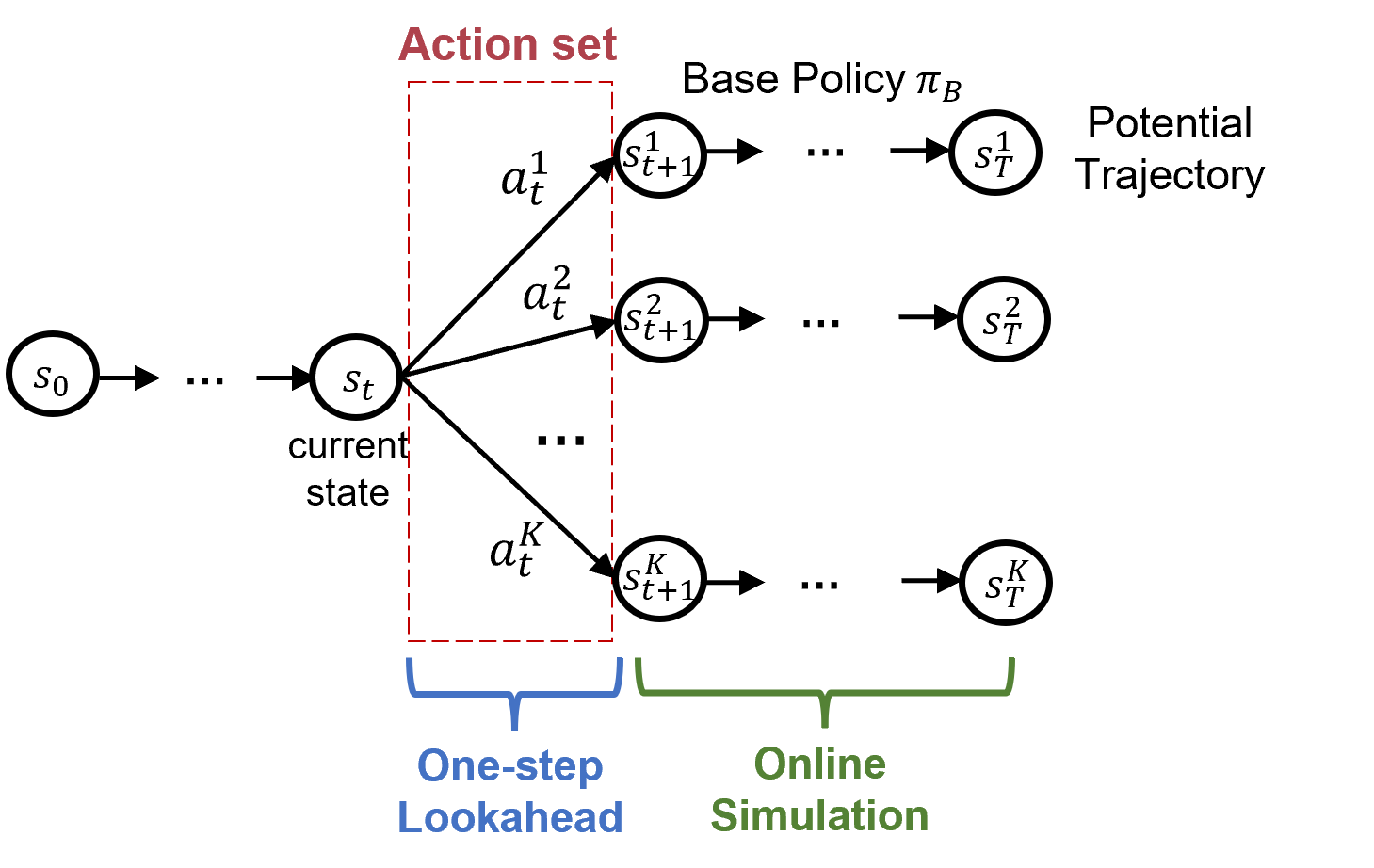}
                \caption{Illustration of Simulation-based ODP}
                \label{fig:ODP_alg}
        \end{figure}
        
        An illustration of the proposed ODP algorithm is presented in Fig. \ref{fig:ODP_alg}. To approximate state-action values efficiently, optimization over the remaining stages in (\ref{eq:Q-fun}) should be avoided. To this end, we employ a base policy $\pi_B(\cdot)$ that maps $(s_t,\xi_t)$ to an action $a_t \in \mathcal A_t$. Then,  the state-action value under the base policy can be approximated as 
        \begin{equation}
        \label{eq:appr-Q-fun}
        Q^\pi(s_t, a_t) =C(s_t, a_t) + \mathbb{E} \left[ \sum_{\tau=t+1}^T C(s_\tau, \pi_B(s_\tau,\xi_\tau))\right]
        \end{equation}
        We use $Q^\pi(s_t, a_t)$ in (\ref{eq:appr-Q-fun}) to approximate $Q(s_t, a_t)$ in (\ref{eq:Q-fun}), so policy optimization over the remaining stages is replaced by the base policy $\pi_B(\cdot)$, and the expectation can be  evaluated by simulation, no optimization problem is to be solved. Indeed, $Q^\pi(s_t, a_t)$ can be good or not, depending on the quality of the base policy $\pi_B$. We leave the selection of base policy to subsection \ref{subsection: base_policy}.

        \subsubsection{Action selection} 
        
        To reduce the search space, a subset of $\mathcal A_t$ containing promising candidate actions $[a_t^1,\cdots,a_t^K]$ is examined. We  denote this subset as $\tilde{A_t}$. For each candidate action $a_t^i \in \tilde{A_t}$, the one-step cost $C(s_t, a_t^i)$ is instantly clear.The selection of $\tilde{A_t}$ is discussed in Section \ref{subsection: action space}.
        
        \subsubsection{Simulation} 
        
        Once $a_t$ is chosen, the system evolves to a new state $s_{t+1}$, and then $\xi_{t+1}$ is observed, based on which the action in period $t+1$ is determined by  $a_{t+1}= \pi_B(s_{t+1},\xi_{t+1})$. This process repeats until the last period $T$. Considering that $\xi_{t+1},\cdots,\xi_T$ are uncertain when evaluating $a_t$, we generate $N_{S}$ scenarios and approximate the expectation in (\ref{eq:appr-Q-fun}) by 
        \begin{equation*}
        Q^\pi(s_t, a^i_t) =C(s_t, a^i_t) + \frac{1}{N_{S}}\sum_{i=1}^{N_S}\sum_{\tau=t+1}^T  C(s_\tau^{(i)}, \pi_B(s_\tau^{(i)}))
        \end{equation*}
        where a scenario represents a possible trajectory of uncertainties $\{\xi_{t+1},\cdots,\xi_T\}$, and $s_{\tau}^{(i)}$ is the system state in period $\tau$ under scenario $i$. 
        
        Finally, the action to be deployed in practice is 
        \begin{equation} 
        \label{eq:action_ODP}
        a_t^* = \mu_t(s_t,\xi_t) = \arg \min_{a_t\in \tilde{A_t}} Q^\pi(s_t, a_t)
        \end{equation}
        This policy is denoted by $\mu_t(s_t,\xi_t)$, and $\mu_S=\{\mu_1,\cdots,\mu_T\}$. Comparing (\ref{eq:Q_action}) and (\ref{eq:action_ODP}), the exact Q-function is replaced with an approximation, potentially making $\mu_t(s_t,\xi_t)$ suboptimal. Nevertheless, policy $\mu_S$ is guaranteed to be superior to the base policy $\pi_B$, as stated below. Empirical observations indicate that this improvement is often substantial.

        \begin{prop}
                Let $\pi_B$ and $\mu_S$ be the base policy and online dynamic programming policy; $J_t^\pi(s_t)$ and $J^\mu_t(s_t)$ are the respective expected cumulative loss function using policy $\pi_B/\mu_S$. The following relation holds:
                \begin{equation} \label{eq:ODP_guarant}
                J_t^{\mu}(s_t) \le J_t^{\pi}(s_t), \ \forall s_t, \forall t
                \end{equation}
        \end{prop}
        
        \begin{proof}
                By Induction. When $t=T$, $J_T^{\mu} = J_T^{\pi}$ as specified by the terminal condition, which is equal to $0$ since all loads have been recovered eventually. Assuming that (\ref{eq:ODP_guarant}) holds for $t=k+1$, that is $J_{k+1}^{\mu}(s_{k+1}) \le J_{k+1}^{\pi}(s_{k+1}), \forall s_{k+1}$, then
                \begin{subequations}
                        \begin{align}
                        & J_k^{\mu}(s_k) \notag \\
                        = &\ C(s_k, \mu_{S}(s_k, \xi_k)) +   \mathbb{E}\left[J_{k+1}^{\mu}(f(s_k, \mu_{S}(s_k), \xi_k)) \right] \label{eq:ODP_proof_1} \\
                        \le &\ C(s_k, \mu_{S}(s_k, \xi_k)) +  \mathbb{E}\left[J_{k+1}^{\pi}(f(s_k, \mu_{S}(s_k), \xi_k)) \right] \label{eq:ODP_proof_2} \\
                        = & \min_{a_k} \left\{C(s_k, a_k) +  \mathbb{E}\left[J_{k+1}^{\pi}(f(s_k, a_k, \xi_k)) \right]\right\} \label{eq:ODP_proof_3} \\
                        \le & \  C(s_k, \pi(s_k, \xi_k)) + \mathbb{E}\left[J_{k+1}^{\pi}(f(s_k, \pi(s_k), \xi_k)) \right] \label{eq:ODP_proof_4} \\
                        = & J_k^{\pi}(s_k)  \notag
                        \end{align}
                \end{subequations}
                where (\ref{eq:ODP_proof_1}) is the recursive equation of $J_k^{\pi_{S}}$; (\ref{eq:ODP_proof_2}) results from the assumption of induction; (\ref{eq:ODP_proof_3}) is an optimization based expression of policy $\mu_S$ in (\ref{eq:action_ODP}), because 
                \begin{equation*}
                Q^\pi(s_k, a_k) = C(s_k, a_k) +  \mathbb{E}\left[J_{k+1}^{\pi}(f(s_k, a_k, \xi_k)) \right]
                \end{equation*}
                Finally,
                (\ref{eq:ODP_proof_4}) is evident since action $a_k$ in (\ref{eq:ODP_proof_3}) is optimal. Based on the principle of induction, (\ref{eq:ODP_guarant}) holds.
        \end{proof}

        Different from standard reinforcement learning which trains Q-function in (\ref{eq:Q-fun}) offline using extensive data and making online decisions via (\ref{eq:Q_action}), the proposed method requires neither historical data nor offline training; all computations are done online. This requires  a solid base policy and reasonable action selection for implementing online simulations, as well as scenario generation methods to handle unknown or imprecise parameters. These factors are discussed in detail in the following subsections.

        \subsection{Base Policy} \label{subsection: base_policy}
        
        According to (\ref{eq:ODP_guarant}), the base policy $\pi_B$ determines the upper bound of the expected cumulative loss, so a good base policy is crucial. Considering the nature of restoration tasks, we adopt priority-based policies for repair crews and mobile emergency generators. Priority rules are simple and leverage human experiences, so can be efficiently simulated and lead to satisfactory strategies.
        
        \subsubsection{Base policy of repair crews} Given the current state $s_t$ of the system in period $t$, repair crews needs to decide the next target lines  $\boldsymbol\chi^R_t=[\beta_{1,t}^R,\cdots,\beta_{N_R,t}^R]$ to repair, where $N_R$ is the number of crews and $\beta_c^R$ is the next target of crew $c$.  Target selection should balance revenue and cost: the former means the amount of recovered load when the fault is fixed, while the latter refers to the time spent on the repairing work. Considering this trade-off, the combined load recovery rate index is defined as the ratio of restored load to the average repair time for lines in the current state $s_t$:
        \begin{equation} 
        \label{eq:CLRR}
        I_{CL}(s_t, \boldsymbol\chi^R_t) = \frac{LV({\Omega^{FL}_t, \boldsymbol\alpha_{t}^M}) - LV({\Omega^{FL}_t/\boldsymbol\chi^R_t}, \boldsymbol\alpha_{t}^M)}{N_R^{-1}\sum_{i=1}^{N_{R}}\left[\mathcal{T}^R(\alpha_{i,t}^{R}, \beta_{i,t}^R) + \mathcal{R}_{\beta_{i,t}^R}\right]} 
        \end{equation}
        where $\boldsymbol\alpha_{t}^M=\{\alpha^{M}_{g,t}, \forall g\in \Omega^M\}$ represents the buses that mobile emergency generators are currently located; $LV({\Omega^{FL}_t, \boldsymbol\alpha_{t}^M})$ is the unsatisfied load under current faults $\Omega^{FL}_t$; $LV({\Omega^{FL}_t/\boldsymbol\chi^R_t}, \boldsymbol\alpha_{t}^M)$ is the unsatisfied load after the faults in $\boldsymbol\chi^R_t$ are all fixed. Both travel time $\mathcal{T}^R(\alpha_{i,t}^{R}, \beta_{i,t}^R)$ and repair time $\mathcal{R}_{\beta_{i,t}^R}$ are considered in the time cost.
        
        This index considers the simultaneous repair of $N_R$ faults by $N_R$ crews, as the impact of each crew's repair work may not be independent. For example, if the restoration of a load requires repairing two upstream lines,  then fixing only one will not restore power. Therefore, the coordination of multiple repairing works is considered in (\ref{eq:CLRR}). Based on this index, the base policy $\pi_B^{RC}$ for repair crews is 
        \begin{equation} \label{eq:BP_RC}
        (\boldsymbol{\chi}^R_t)^{\pi_B^{RC}} = \argmax_{\boldsymbol\chi^R \subseteq \Omega^{FL}_t} \ I_{CL}(s_t, \boldsymbol\chi^R)
        \end{equation}
        This implies that repair crews should select the target line combination that maximizes load recovery efficiency.
        
        Another issue is how to efficiently calculate the load loss value $LV$ in (\ref{eq:CLRR}). Since the simulation yields an approximate solution, solving an exact optimal power flow to compute $LV$ may be unnecessary. A method for estimating $LV$ will be introduced later.

        \subsubsection{Base policy of mobile emergency generators} 
        
        Given the current state $s_t$ of the system in period $t$, mobile emergency generators need to choose target buses $\boldsymbol\chi^M_t =[\beta_{1,t}^M,\cdots,\beta_{N_M, t}^M]$ for power supply. Because of time and fuel consumption during traveling,  mobile emergency generators are inclined to stay at a specific bus. A reasonable index to generate better connection buses for mobile generators is
        \begin{equation} \label{eq:IDL}
        I_{DL}(s_t, \boldsymbol\chi^M_t) = LV({\Omega^{FL}_t, \boldsymbol\alpha_{t}^M}) - LV({\Omega^{FL}_t, {\boldsymbol\chi^M_t}})
        \end{equation}
        Given the current fault lines $\Omega^{FL}_t$, $LV({\Omega^{FL}_t, \boldsymbol\chi^M_t}) $ represents unsatisfied load when mobile generators are positioned at buses ${\boldsymbol\chi^M_t}$;  Index (\ref{eq:IDL}) quantifies the change in unsatisfied load when mobile generators relocate from $\boldsymbol \alpha_{t}^M$ to $\boldsymbol\chi^M_t$.  The base policy $\pi_B^{MEG}$ of  mobile generators is then determined from
        \begin{equation} \label{eq:BP_MEG}
        (\boldsymbol{\chi}^M_t)^{\pi_B^{MEG}} = \argmax_{\boldsymbol\chi^M \subseteq \Omega^{AP}} \  I_{DL}(s_t, \boldsymbol\chi^M)\\
        \end{equation}
        indicating that mobile emergency generators will move to buses where they can maximize power supply.
        
        \subsubsection{Base policy of energy storage and line switches} In addition to the mobile resources mentioned, the base policy should also account for the operation of fixed flexible resources including energy storage and line switches.
        
        For energy storage, the policy is as follows: At each time step, calculate the maximum output of other generation resources on the island. If this output surpasses the island's demand, store the surplus energy. Otherwise, discharge stored power to provide emergency supply. "The discharge power is determined by dividing the available stored energy by the estimated outage duration of the island's most critical loads (i.e., those with the highest $c_i$). The outage duration is estimated via simulation within the ODP algorithm.

        For line switches, the following policy is used to maintain the radial topology of the system during simulation: At each time step, use a depth-first search algorithm to detect any loops in the network. If a loop is detected, disconnect the line switch that minimizes load loss variation. As high accuracy is not required for future power flow simulation, this approach provides a fast and practical method for enforcing the network's radial topology.

        \subsubsection{Simulation of future load loss} Using the proposed base policies, the actions of various emergency resources can be determined during simulations. By incorporating the state transition function (\ref{State_Trans}), the simulated future state $s_\tau$ is obtained. Furthermore, we need to evaluate  $C(s_\tau, a_\tau), \forall \tau>t$ during the simulation, which requires power flow simulation. Exact computation entails minimizing a loss function subject to constraints (\ref{eq:DN_power})(\ref{eq:DN_radial})(\ref{eq:MEG_power}). However, in online simulation, efficiency is the primary concern, and approximation is acceptable. Here we suggest an efficient way to estimate load loss in simulation:
        \begin{itemize}
                \item Based on the fault status of each distribution line and the state of line switches as determined by the base policy, partition the system into isolated islands and assign an island index to each bus.
                
                \item For each island, compute the active power supply, including that from mobile generators and energy storage acquired by the base policy. Allocate the power to each load according to its priority, based on its weight $c_i$.
                
                \item Determine the load loss $P_{i,t}^D-p_{i,t}^D$ at each bus $i$ according to the allocation of power resources.
        \end{itemize}
        Notably, this procedure approximates future losses $C(s_\tau, a_\tau)$ for $\tau = t+1,\cdots, T$. Conversely, the real-time load loss $C(s_t, a_t)$ in the current period $t$, along with action $a_t^{DN}$, should be accurately obtained from the optimal power flow, which will be discussed in Section \ref{subsection: DN_alg}.
        
        \subsection{Selecting Promising Actions} \label{subsection: action space}
        
        In view of the potential collaboration of multiple mobile resources, the joint scheduling of repair crews and mobile generators involves a high-dimensional action space. To reduce the computation burden of online simulation, the values of $Q^\pi(s_t, a_t)$ over a small subset $\tilde{A_t}$ that consists of several promising actions are evaluated.

        Instead of jointly scheduling all mobile resources, we optimize the dispatch of repair crews and mobile emergency generators in sequence. The motivation is that establishing power supply pathways has a long-term effect. Once repair action is made, mobile emergency generators go to those islands that cannot acquire power supply temporarily.
        
        Leveraging insights from the proposed base policy, we use the metrics $I_{CL}$ in (\ref{eq:CLRR}) and $I_{DL}$ in (\ref{eq:BP_MEG}) to reduce the action space. For the repair crew action $\bm{\chi}_t^R$, assume that all possible repair target combinations form the original action set $A_t^{RC}$. For each action $\left(\bm{\chi}_t^R\right)^j$ in $A_t^{RC}$, compute the corresponding index $I_{CL}\left(\left(\bm{\chi}_t^R\right)^j\right)$. Let $top_{K_a}$ represent the operation that returns the indices of the $K_a$ largest elements in a set. The reduced action set for repair crews is then: 
        
        \begin{equation}
        \tilde{A}_t^{R C}=\left\{\left(\bm{\chi}_t^R\right)^{j_1^R}, \ldots,\left(\bm{\chi}_t^R\right)^{j_{K_a}^R}\right\}
        \end{equation}
        where $\left\{j_1^R, \ldots, j_{K_a}^R\right\} = \operatorname{top}_{K_a}\left(I_{CL}\left(\bm{\chi}^i\right)\right)$, $\forall \bm{\chi}^i \in A_t^{RC}$
        
        Similarly, form the reduced action set $\tilde{A}_t^{MEG}$ for the action of mobile generators using the $top_{K_a}$ values of $I_{DL}$.

        \subsection{Scenario Generation} \label{subsection: scenario}
        
        To approximate expectation in (\ref{eq:appr-Q-fun}), $M_s$ scenarios are generated to mimic uncertainty. Based on existing research \cite{pdf_repair_time, pdf_solar, Decomposed_SP_1}, the prior distribution of repair times follows a Weibull distribution, the forecast error of solar output follows a normal distribution, and the forecast error of load curves follows a uniform distribution. Other uncertain factors can also be sampled according to their respective distributions. The simulations across different scenarios can be run in parallel, so computational efficiency is not a significant obstacle to the implementation of the proposed method.

        \subsection{Dispatch of the Distribution Network}\label{subsection: DN_alg}
        
        Based on the previous discussion, the dispatch decisions of repair crews $a_t^{RC}$ and mobile emergency generators $a_t^{MEG}$ can be successively obtained by the proposed ODP algorithm, which will, in turn, influence the system state. Meanwhile, the real-time power flow decision $a_t^{DN}$ and the current loss $C(s_t, a_t)$ are obtained by the following optimal power flow problem:
        
        \begin{equation} \label{eq:single_OPF}
                \begin{aligned}
                C(s_t,a_t)=\min~~ & \sum_{i=1}^{N_B} c_i (P_{i,t}^D-p_{i,t}^D) \Delta t \\
                \mbox{s.t.}~~&  (\ref{eq:DN_power}), (\ref{eq:DN_radial}), (\ref{eq:MEG_power})
                \end{aligned}
        \end{equation}
        
        Combining the details in Section \ref{section: solution}, the online decision procedure for distribution network restoration is summarized in Algorithm \ref{alg:online}; in particular, the ODP algorithm is given in Algorithm \ref{alg:ODP}.
        
        \begin{algorithm}[htbp]
                \caption{Online Restoration}
                \label{alg:online}
                \begin{algorithmic}[1] 
                        \State Initialize system state $s_0$
                        \For{$t = 1,\cdots,T$}
                        \State Observe $\xi_{t}$
                        \State For repair crews and mobile generators, update target lines and target buses by \newline \hspace*{1.1em} $[\boldsymbol{\chi}^R_t$, $\boldsymbol{\chi}^M_t]= \mbox{ODP}(s_t, \xi_t, K_a)$
                        \State Optimize power flow decision $a_t^{DN}$ by (\ref{eq:single_OPF}), and the  optimal loss $C(s_t, a_t)$  .
                        \If{$\Omega^{FL}_t = \emptyset$}
                        \State break
                        \EndIf
                        \State Update state $s_{t+1} = f(s_{t}, a_{t}, \xi_{t})$
                        \EndFor
                        \State Acquire total load loss value $\sum_{t=1}^T C(s_t, a_t)$
                \end{algorithmic}
        \end{algorithm}
        
        \begin{algorithm}[htbp]
                \caption{ODP algorithm}
                \label{alg:ODP}
                \begin{algorithmic}[1]
                        \Statex \textbf{Input:} state $s_t$; observation $\xi_t$; action set size $K_a$
                        \Statex \textbf{Output:} optimal target of repair crews $\boldsymbol{\chi}^{R*}_t$, optimal target of mobile emergency generators $\boldsymbol{\chi}^{M*}_t$
                        \State Generate $N_{S}$ scenarios based on the distribution of unknown repair time and renewable output.
                        \State Based on the observed faults in $\Omega_t^{FL}$ and $\Omega_t^{NL}$, construct the reduced action set $\tilde{A}_t^{RC}=\{(\boldsymbol{\chi}^R_t)^1,\cdots,(\boldsymbol{\chi}^R_t)^{K_a}\}$ using the combined load recovery rate index index (\ref{eq:CLRR}).
                        \State Extract the optimal target of repair crews $\boldsymbol{\chi}^{R*}_t$ by (\ref{eq:action_ODP}).
                        \State Based on the accessible points in $\Omega^{AP}$, construct the reduced action set $\tilde{A}_t^{MEG}=\{(\boldsymbol{\chi}^M_t)^1,\cdots,(\boldsymbol{\chi}^M_t)^{K_a}\}$ using the load loss variation index (\ref{eq:BP_MEG}).
                        \State Obtain the optimal target of mobile emergency generators $\boldsymbol{\chi}^{M*}_t$ by (\ref{eq:action_ODP}).
                \end{algorithmic}
        \end{algorithm}

        \section{Case Study} \label{section: case study}
        The proposed method is implemented in Python 3.10 and tested on distribution systems with 123 and 8500 buses. All experiments are performed on a laptop with Intel Core Ultra 5 125H CPU and 16 GB of memory.

        \subsection{123-Bus System}
\begin{figure}[!htbp]
                \centering
                \includegraphics[width=0.85\linewidth]{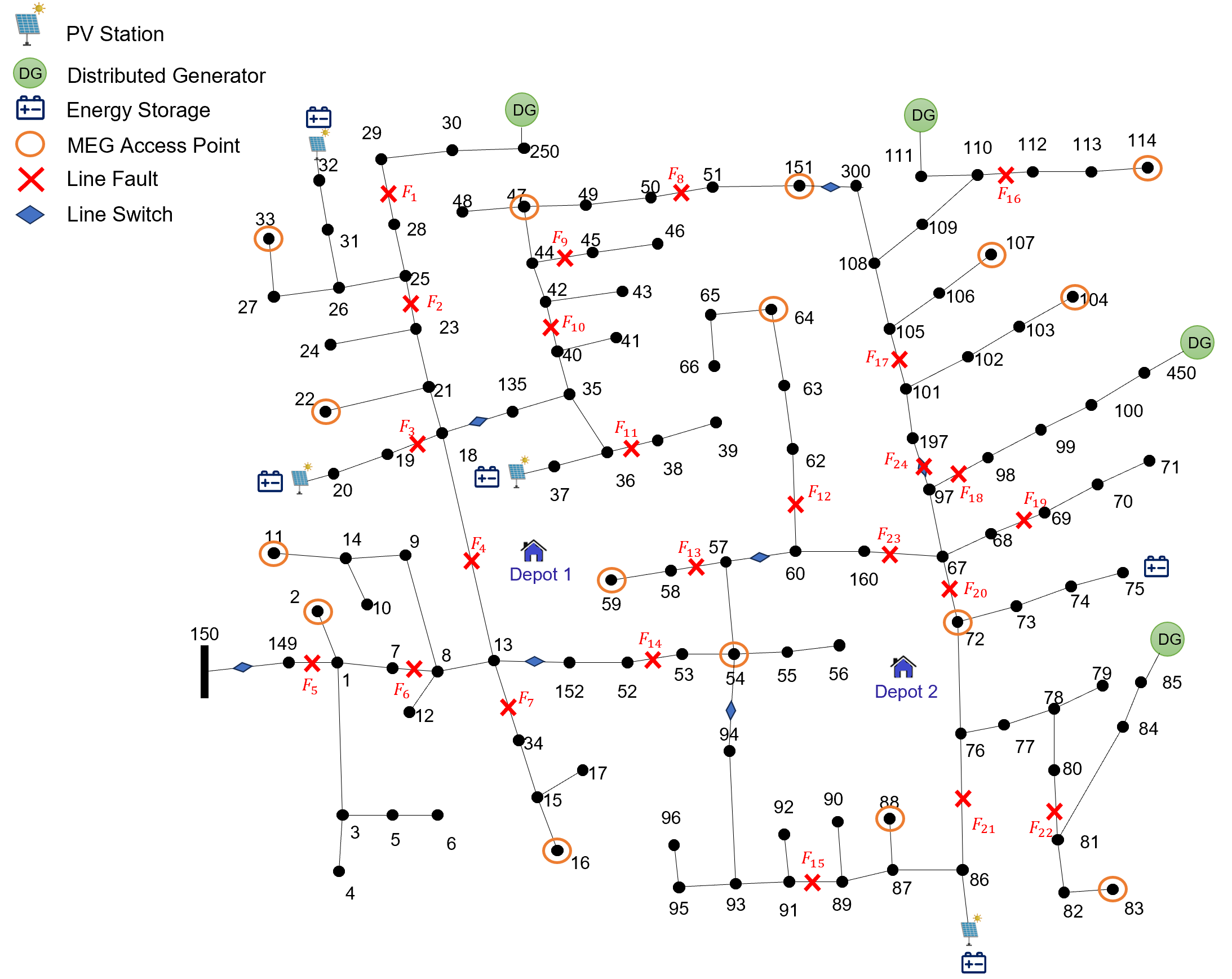}
                \caption{123-bus system}
                \label{fig:case123}
                \vspace{-5pt}
        \end{figure}

\begin{figure}[!h]
                \centering
                \includegraphics[width=0.5\linewidth]{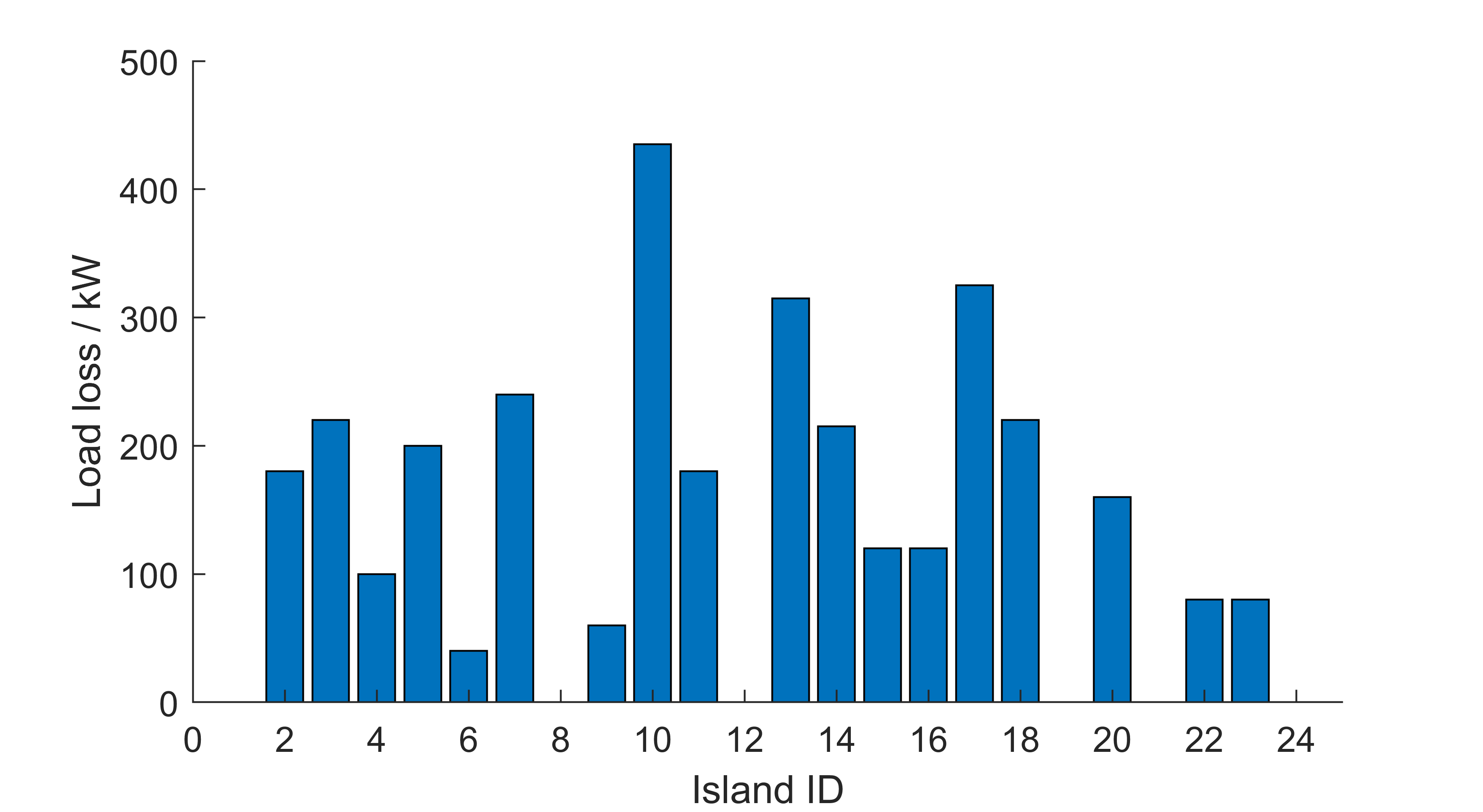}
                \caption{load loss of each island}
                \label{fig:isl_loss_123}
        \end{figure}

\begin{figure}[!h]
                \centering
                \includegraphics[width=0.8\linewidth]{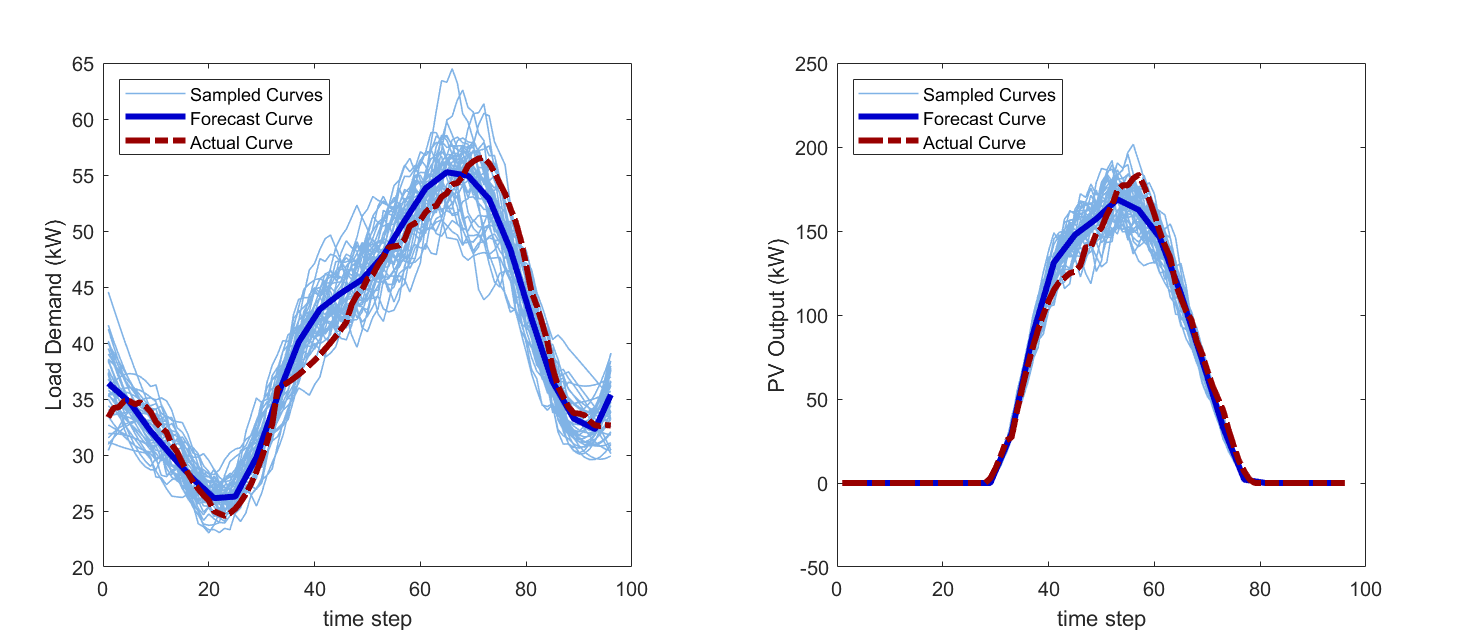}
                \caption{Example of scenarios of load demand and photovoltaic output}
                \label{fig:scenario}
        \end{figure}

        The 123-bus system is shown in Fig. \ref{fig:case123}. The system includes four identical micro gas turbines (total capacity: 1.6 MW), four photovoltaic stations (total capacity: 800 kW), and five energy storage units (total capacity: 300 kW/600 kWh). There are two depots in the system, each with two repair crews and one mobile emerging generator.  Travel time in the nominal condition is proportional to the Manhattan distance between the origin and the destination. Following an extreme event, 24 transmission lines are damaged, dividing the 123-bus system into 25 isolated islands. The initial load loss for each island is illustrated in Fig. \ref{fig:isl_loss_123}. Load shedding costs $c_i$ range from $5\$/$kWh to $20\$/$kWh. Repair times for faulty lines are randomly generated within a range of 30 minutes to 2 hours. Each restoration period lasts $\Delta t=15$ minutes. Faults $F_1, F_4,$ and $ F_{21}$ are discovered in periods $4, 6, 12$ respectively, while all other faults are detected at $t=0$. An example of generated scenarios of load demand and photovoltaic output, are displayed in Fig. \ref{fig:scenario}. Complete system data are available in \cite{data_github}.
        
        The following online decision-making approaches are compared in the case studies:
        
        \textbf{ODP}: During each online decision update in the restoration process, the ODP algorithm first identifies 40 promising candidate actions according to the method in Section \ref{subsection: action space}. Their Q-values are then estimated through online simulations with 50 scenarios, utilizing the base policy described in Section \ref{subsection: base_policy}.
        
        \textbf{Base Policy}: Direct application of the priority list-based policy developed in Section \ref{subsection: base_policy}. The base policy does not involve any optimization for the restoration strategy, and therefore can be used as a baseline for comparison with other methods.
        
        \textbf{MPC}: Model predictive control solves a $H$-step lookahead version of MIP model (\ref{eq:opt_restore_MIP}) in Section \ref{section:MIP}. At each time step, a truncated MIP problem is formulated with decision variables and constraints for the next $H$ steps and solved using the GUROBI solver. Only the first-step decision is implemented, and the optimization is re-evaluated in subsequent steps as new information becomes available. While repair crew routing variables do not have an explicit time dimension, the routes within the next $H$ steps are constrained by the arrival time constraints (\ref{eq:RC_recovery_2})-(\ref{eq:RC_recovery_3}) in the MPC problem, with only the first repair target being executed. Given the problem scale, the lookahead step size $H$ is set to 12, allowing the solver to find a meaningful solution within 10 minutes, which aligns with the 15-minute decision update interval. Uncertain parameters, such as repair time and renewable output, are approximated using their mean values for prediction.

        \textbf{Two-stage SP}: In two-stage stochastic programming (SP), the first-stage decision involves pre-determining a comprehensive routing plan for mobile resources while accouting for uncertainty scenarios regarding repair time, renewable output, and load demand. Since these decisions are made before uncertainty unfolds, the routing schedules are pre-planned before the start of the restoration process (i.e., once most damages have been assessed after the disaster), and remain unchanged throughout the restoration. If new faults are discovered during restoration, they are directly appended to the end of the pre-planned schedule without modifying the existing routing. The second-stage decision, on the other hand, consists of real-time adjustments for power dispatch, topology reconfiguration, and load supply after uncertainty is realized. While two-stage SP accounts for uncertainties when making first-stage decisions, it follows a fixed, pre-planned routing for mobile resources without subsequent online adjustments.
        
\begin{figure}[h]
                \centering
                \includegraphics[width=0.6\linewidth]{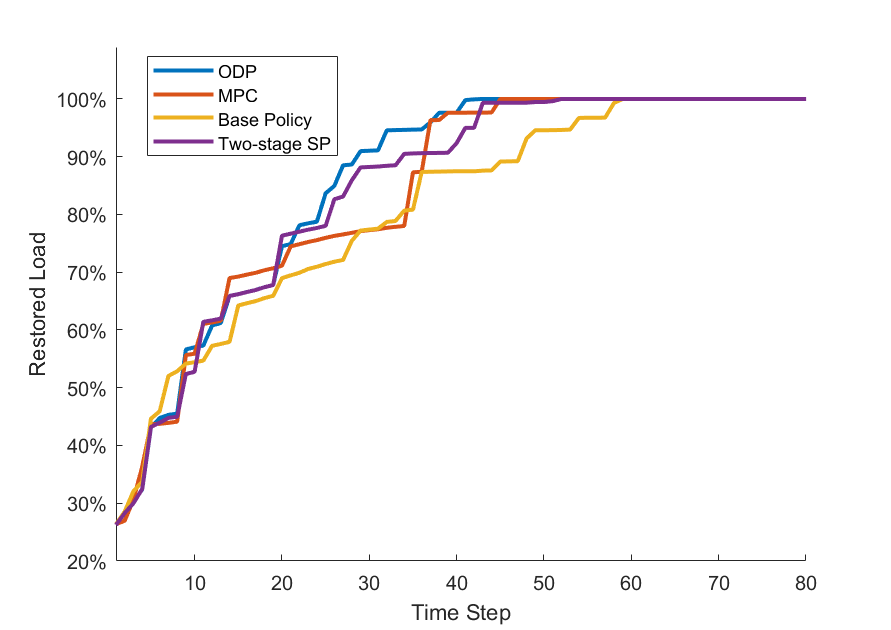}
                \caption{Load restoration for 123-bus case}
                \label{fig:WLL}
        \end{figure}

        Results of different methods are compared in Table \ref{table:123_result_comp} and Fig. \ref{fig:WLL}. In Table, the online computation time refers to the maximum computational time required for decision updates in a single time period, while performance improvement indicates the percentage decrease in the objective value compared to the base policy, reflecting the extent of optimization in the restoration strategy.  It is worth noting that, in two-stage SP, the mobile resource routing plan is pre-determined in the first stage rather than being updated in real time during the second stage. As a result, there is no corresponding online computation time in Table \ref{table:123_result_comp}.
        
        \begin{table}[h]
                \centering
                \scriptsize
                \renewcommand{\arraystretch}{1.3}
                \caption{Results of each method on the 123-bus case}  \label{table:123_result_comp}     
                \begin{tabular}{c|ccc}
                        \toprule
                        & Objective value  & Online computation time & Performance improvement  \\
                        \midrule
                        Base Policy &  \$176,235 & 0.11s & -  \\
                        ODP    & \$121,601 & 89s & 31.0\%  \\
                        MPC & \$147,782  & 573s & 14.9\% \\
                        Two-stage SP & \$132,633  & - & 24.7\% \\
                        \hline
                \end{tabular}
                
        \end{table}
        
       As shown in Table \ref{table:123_result_comp}, ODP outperforms all other methods in reducing load loss, achieving a 31\% improvement over the base policy. Moreover, its online computation time remains under 2 minutes, ensuring efficient decision updates. In contrast, while MPC also updates decisions online, it requires about 10 minutes per update due to the complexity of solving the MIP, limiting its practicality for real-time applications. Given the need for decision updates every 15 minutes, MPC can only plan up to 12 time steps ahead, restricting its ability to incorporate distant restoration targets and reducing its overall effectiveness. ODP, on the other hand, relies solely on algebraic computations during simulation, enabling faster decision-making and allowing for a more comprehensive evaluation of long-term impacts across different scenarios. This efficiency and foresight contribute to its significantly superior performance.

       Table \ref{table:123_result_comp} primarily presents the models' performance based on the objective value. To further highlight the superiority of the proposed ODP method in terms of restoration performance, Table \ref{table:123_result_index} compares additional resilience indices. In addition to the objective of economic loss (EL), two other indices are also considered: Total Energy Not Served (TENS), which represents the total unserved load demand, and Weighted Average Outage Duration (WAOD), which reflects the system's weighted average outage duration, with the load shedding cost at each bus as the weight. As shown in Table \ref{table:123_result_index}, the ODP method outperforms all other methods across these indices, demonstrating its effectiveness.

       \begin{table}[h] 
           \centering 
           \scriptsize 
           \renewcommand{\arraystretch}{1.3} 
           \caption{Results with Different Resilience Indices} \label{table:123_result_index}
           \begin{tabular}{c|ccc} 
               \toprule
               & EL (\$) & TENS (kWh) & WAOD (hour) \\
               \midrule 
               Base Policy & 176,235 & 15,883 & 6.65 \\
               ODP & 121,601 & 10,695 & 3.13 \\
               MPC & 147,782 & 14,615 & 4.68 \\
               Two-stage SP & 132,633 & 11,980 & 3.56 \\
               \hline 
           \end{tabular} 
       \end{table}

        Routing results of mobile resources in different methods are illustrated in Table \ref{table:route_comp_123}, from which the superiority of ODP over MPC can be observed. The proposed ODP method first dispatches $RC_1$ and $RC_2$ to repair $F_5$ and $F_6$, thereby restoring the connection between the system and substation bus 150. This schedule adopts a relatively global perspective. On the other hand, MPC prioritizes sending $RC_1$ and $RC_2$ to repair $F_3$ and $F_8$, aiming to maximize the emergency supply capabilities of the photovoltaic and energy storage systems at nodes 20 and 37 in Fig. \ref{fig:case123}. However, considering the fluctuation of the photovoltaic systems and the energy constraints of the storage units, these resources lack the capacity for sustained long-term output, and the MPC strategy is less optimal than ODP. The primary reason lies in the limited forecasting horizon of MPC, which only extends three hours ahead. In contrast, ODP can more quickly and effectively simulate the system's global trajectory, making it more suitable for online decision-making.

   Compared to ODP, the scheduling plan of Two-stage SP, as presented in Table \ref{table:route_comp_123}, initially exhibits some similarities but diverges in later stages. This difference stems from ODP's dynamic adaptability, allowing real-time adjustments based on newly discovered faults and observed repair times. In contrast, the dispatching plan of Two-stage SP is largely predetermined, lacking real-time adjustments for emerging faults like F1 and F4 or variations in repair durations. For instance, $RC_1$'s repair of F14 takes nearly two hours, significantly longer than expected. Unlike two-stage SP, ODP dynamically reallocates subsequent repair plans in real time, enhancing  flexibility and efficiency.

        Additionally, Table \ref{table:route_comp_123} also reveals significant differences in the mobile resource routing strategies between ODP and the base policy. While ODP leverages the base policy for future simulations, it optimizes current decisions by accounting for the long-term impact of various candidate actions, overcoming the myopic nature of the base policy. This enables ODP to achieve superior performance, unaffected by the base policy's limitations, offering an advanced approach that balances solution quality with computational efficiency.

\begin{table}[!htbp]
    \centering
    \scriptsize
    \renewcommand{\arraystretch}{1.5}
    \caption{The routing of mobile resources using different methods} \label{table:route_comp_123}   
    \begin{tabular}{c|l|c|l}
        \toprule
        \multicolumn{2}{c|}{\textbf{ODP}} & \multicolumn{2}{c}{\textbf{MPC}} \\
        \midrule
        $RC_1$ & depot\ 1 $\rightarrow$ F$_5$ $\rightarrow$ F$_{14}$ $\rightarrow$ F$_{11}$ $\rightarrow$ F$_2$ $\rightarrow$ F$_1$ $\rightarrow$ F$_8$ & 
        $RC_1$ & depot\ 1 $\rightarrow$ F$_3$ $\rightarrow$ F$_4$ $\rightarrow$ F$_6$ $\rightarrow$ F$_7$ $\rightarrow$ F$_5$ $\rightarrow$ F$_{15}$ \\
        
        $RC_2$ & depot\ 1 $\rightarrow$ F$_6$ $\rightarrow$ F$_7$ $\rightarrow$ F$_4$ $\rightarrow$ F$_{10}$ $\rightarrow$ F$_9$ $\rightarrow$ F$_3$ &  
        $RC_2$ & depot\ 1 $\rightarrow$ F$_8$ $\rightarrow$ F$_9$ $\rightarrow$ F$_1$ $\rightarrow$ F$_2$ $\rightarrow$ F$_{10}$ $\rightarrow$ F$_{16}$ \\
        
        $RC_3$ & depot\ 2 $\rightarrow$ F$_{15}$ $\rightarrow$ F$_{13}$ $\rightarrow$ F$_{17}$ $\rightarrow$ F$_{16}$ $\rightarrow$ F$_{12}$ $\rightarrow$ F$_{21}$ &  
        $RC_3$ & depot\ 2 $\rightarrow$ F$_{19}$ $\rightarrow$ F$_{18}$ $\rightarrow$ F$_{13}$ $\rightarrow$ F$_{11}$ $\rightarrow$ F$_{17}$ $\rightarrow$ F$_{21}$ \\
        
        $RC_4$ & depot\ 2 $\rightarrow$ F$_{19}$ $\rightarrow$ F$_{18}$ $\rightarrow$ F$_{20}$ $\rightarrow$ F$_{23}$ $\rightarrow$ F$_{24}$ $\rightarrow$ F$_{22}$ &  
        $RC_4$ & depot\ 2 $\rightarrow$ F$_{23}$ $\rightarrow$ F$_{20}$ $\rightarrow$ F$_{24}$ $\rightarrow$ F$_{12}$ $\rightarrow$ F$_{14}$ $\rightarrow$ F$_{22}$ \\
        
        $MEG_1$ & depot\ 1 $\rightarrow$ bus\ 47 $\rightarrow$ bus\ 64 &  
        $MEG_1$ & depot\ 1 $\rightarrow$ bus\ 59 $\rightarrow$ bus\ 114 \\
        
        $MEG_2$ & depot\ 2 $\rightarrow$ bus\ 72 $\rightarrow$ bus\ 114 &  
        $MEG_2$ & depot\ 2 $\rightarrow$ bus\ 54 $\rightarrow$ bus\ 88 \\
        
        \midrule
        \multicolumn{2}{c|}{\textbf{Base Policy}} & \multicolumn{2}{c}{\textbf{Two-stage SP}} \\
        \midrule
        $RC_1$ & depot\ 1 $\rightarrow$ F$_1$ $\rightarrow$ F$_5$ $\rightarrow$ F$_6$ $\rightarrow$ F$_7$ $\rightarrow$ F$_{21}$ $\rightarrow$ F$_{17}$ &  
        $RC_1$ & depot\ 1 $\rightarrow$ $F_5$ $\rightarrow$ F$_{14}$ $\rightarrow$ F$_{10}$ $\rightarrow$ F$_9$ $\rightarrow$ F$_3$ $\rightarrow$ F$_4$ \\
        
        $RC_2$ & depot\ 1 $\rightarrow$ F$_{14}$ $\rightarrow$ F$_{13}$ $\rightarrow$ F$_{18}$ $\rightarrow$ F$_{15}$ $\rightarrow$ F$_{23}$ &  
        $RC_2$ & depot\ 1 $\rightarrow$ F$_6$ $\rightarrow$ F$_7$ $\rightarrow$ F$_{13}$ $\rightarrow$ F$_{11}$ $\rightarrow$ F$_2$ $\rightarrow$ F$_1$ \\
        
        $RC_3$ & depot\ 2 $\rightarrow$ F$_{20}$ $\rightarrow$ F$_{19}$ $\rightarrow$ F$_{16}$ $\rightarrow$ F$_8$ $\rightarrow$ F$_4$ $\rightarrow$ F$_3$ $\rightarrow$ F$_2$ &  
        $RC_3$ & depot\ 2 $\rightarrow$ F$_{19}$ $\rightarrow$ F$_{18}$ $\rightarrow$ F$_{20}$ $\rightarrow$ F$_{22}$ $\rightarrow$ F$_{15}$ $\rightarrow$ F$_{21}$ \\
        
        $RC_4$ & depot\ 2 $\rightarrow$ F$_{22}$ $\rightarrow$ F$_{12}$ $\rightarrow$ F$_{24}$ $\rightarrow$ F$_9$ $\rightarrow$ F$_{11}$ $\rightarrow$ F$_{10}$ &  
        $RC_4$ & depot\ 2 $\rightarrow$ F$_{23}$ $\rightarrow$ F$_{17}$ $\rightarrow$ F$_{16}$ $\rightarrow$ F$_{24}$ $\rightarrow$ F$_{12}$ $\rightarrow$ F$_8$ \\
        
        $MEG_1$ & depot\ 1 $\rightarrow$ bus\ 47 &  
        $MEG_1$ & depot\ 1 $\rightarrow$ bus\ 47 $\rightarrow$ bus\ 114 \\
        
        $MEG_2$ & depot\ 2 $\rightarrow$ bus\ 54 $\rightarrow$ bus\ 72 &  
        $MEG_2$ & depot\ 2 $\rightarrow$ bus\ 72 $\rightarrow$ bus\ 54 \\
        
        \bottomrule
    \end{tabular}
\end{table}

\begin{figure}[h]
    \centering
    \includegraphics[width=1.0\linewidth]{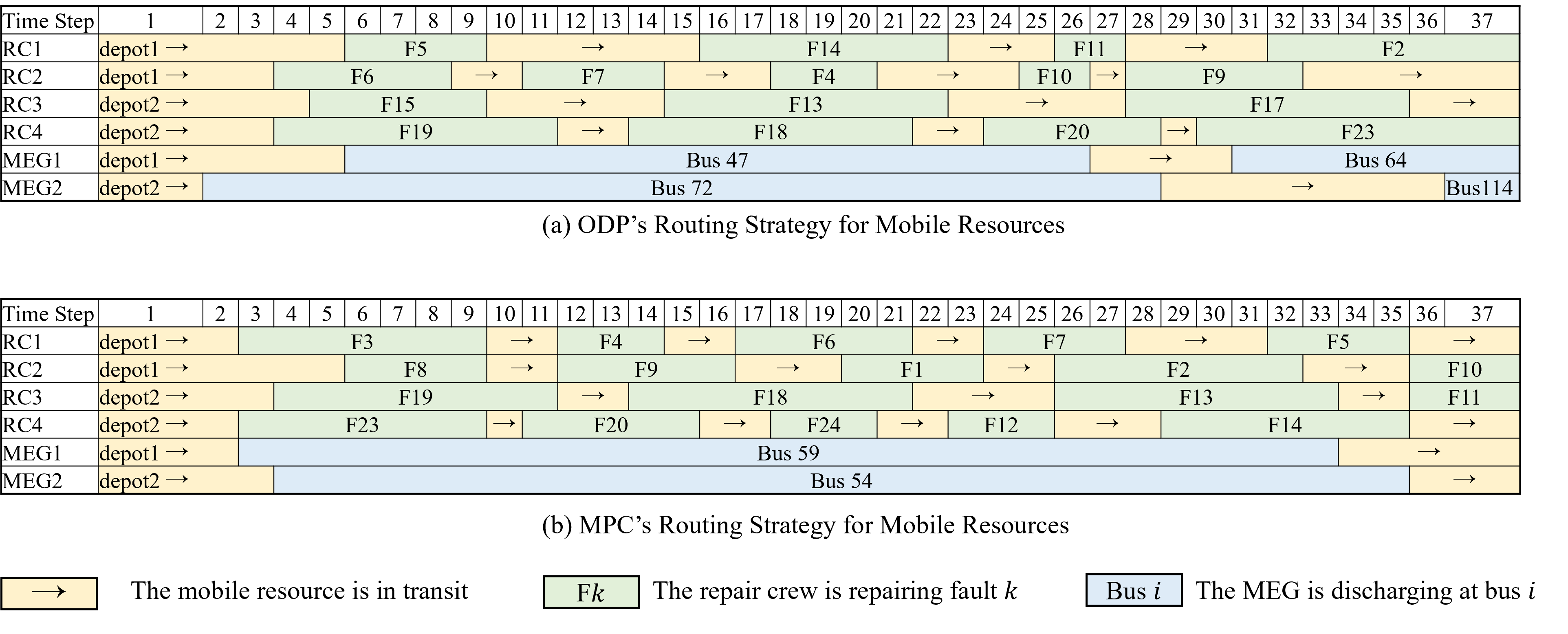}
    \caption{Restoration scheduling of ODP and MPC prior to time step 37}
    \label{fig:time_table_comp}
\end{figure}

\begin{figure}[!h]
    \centering
    \includegraphics[width=1.0\linewidth]{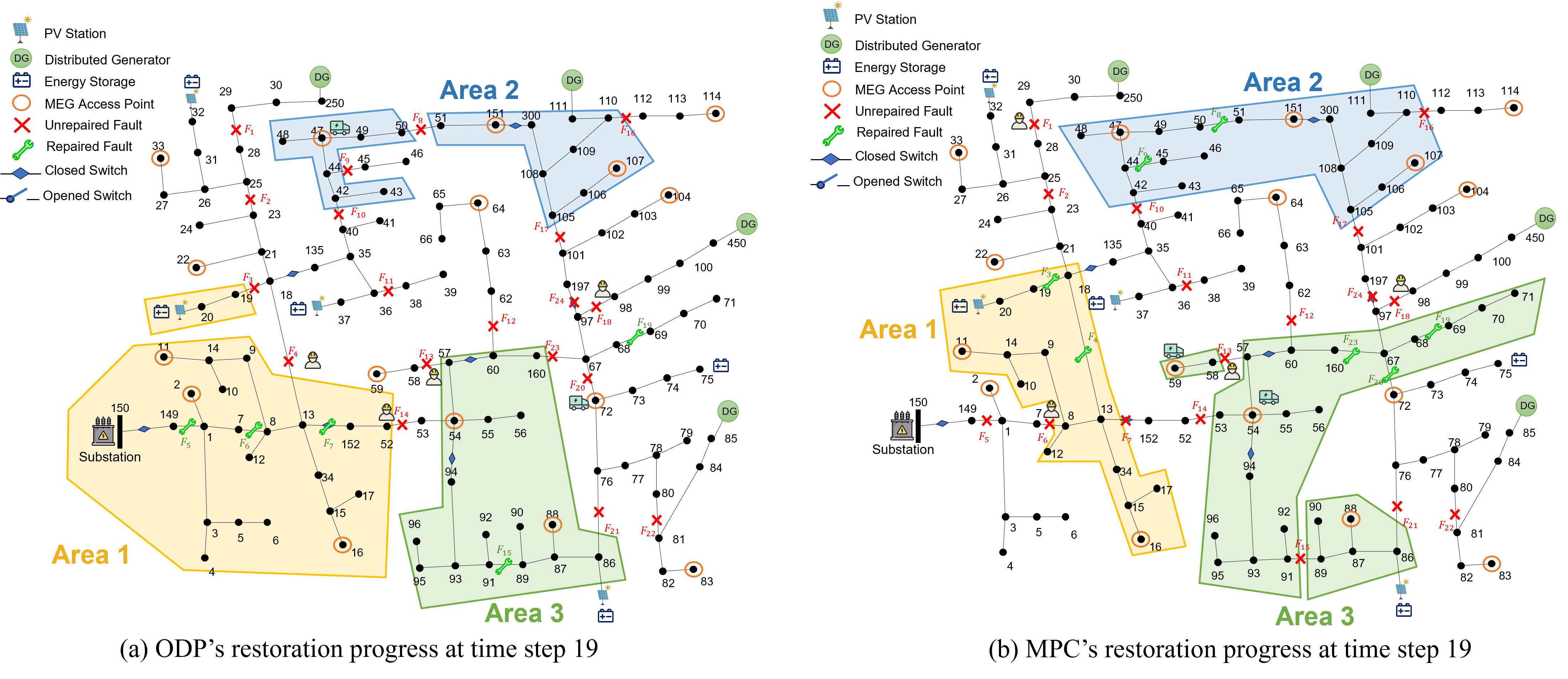}
    \caption{Restoration progress of ODP and MPC at time step 19}
    \vspace{-10pt}
    \label{fig:progress_19}
\end{figure}

\begin{figure}[!h]
    \centering
    \includegraphics[width=1.0\linewidth]{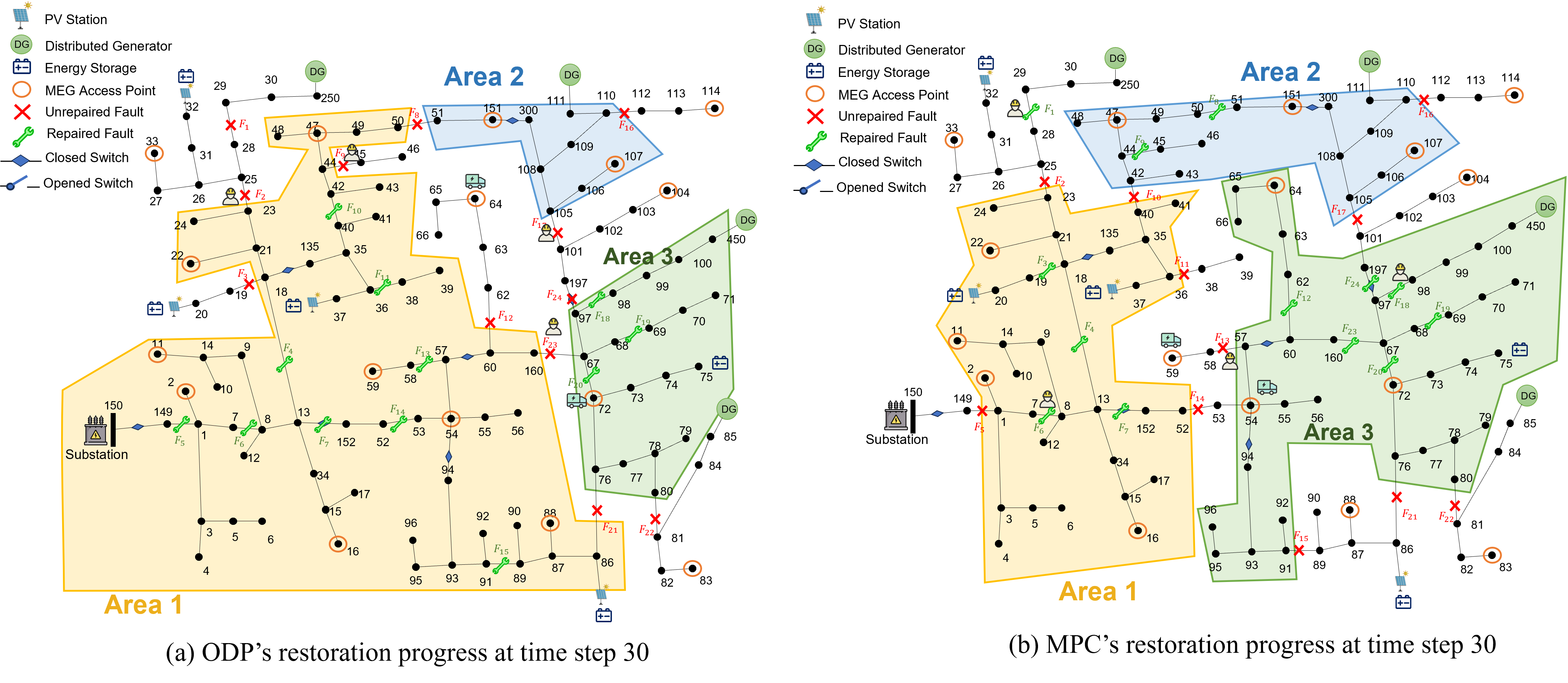}
    \caption{Restoration progress of ODP and MPC at time step 30}
    \vspace{-10pt}
    \label{fig:progress_30}
\end{figure}

It is noteworthy that in Fig. \ref{fig:WLL}, the superiority of ODP becomes particularly evident in the later stages (time steps 20-40), especially around time step 30, where it achieves significantly higher load restoration than other methods, particularly MPC and the base policy. To gain deeper insight into this phenomenon, Fig. \ref{fig:time_table_comp} presents the detailed scheduling of mobile resources under ODP and MPC, with a focus on two key aspects: why load restoration levels are comparable at time step 19 and how ODP achieves a distinct advantage by time step 30. Based on the detailed schedules in Fig. \ref{fig:time_table_comp}, the restoration progress of ODP and MPC at time steps 19 and 30 is further depicted in Fig. \ref{fig:progress_19} and Fig. \ref{fig:progress_30}, respectively.

First, based on Fig. \ref{fig:time_table_comp}, we analyze why the load restoration levels of ODP and MPC show minimal differences at time step 19. As shown in Fig. \ref{fig:time_table_comp}, at this stage, the impact of different restoration strategies is primarily concentrated in three areas. Compared to MPC, the proposed ODP method achieves additional load restoration in Area 1 by restoring  buses 1-7 and enhancing the connectivity to the substation. However, in Area 3, ODP restores less load than MPC, particularly for buses 67-71. In Area 2, the overall restoration effectiveness of both methods is comparable, though their approaches differ: ODP restores the load at buses 42-50 through MEG integration, whereas MPC restores the same area by repairing fault F8, enabling the distributed generation at bus 111 to supply power. Overall, at time step 19, either method can only achieve local load restoration. While ODP restores more load in Area 1 and MPC restores more in Area 3, the total load restoration levels of these two methods remain similar.

However, the schedule in the early-stage not only affects local restoration at that time but also influences subsequent schedules and restoration progress, as evidenced by the restoration status at time step 30 in Fig. \ref{fig:progress_30}. In Fig. \ref{fig:progress_30}, ODP benefits from its early restoration decisions---such as the proactive repair of faults F5-F7---allowing most of Area 1 to be interconnected and supplied by the substation. Given the fluctuations in both bus loads and photovoltaic generation, this interconnectivity enhances the complementarity between power sources and loads, significantly improving the overall restoration level. Meanwhile, under ODP's strategy, only small sections of Areas 2 and 3 require emergency power from distributed generation and MEG, reducing reliance on distributed generation output. In contrast, MPC heavily depends on distributed generation for large-scale emergency supply in Area 1-3, which is inherently limited in capacity. Due to MPC's short-sightedness, it relies more on the emergency supply of distributed generation in the early stages rather than restoring critical faults such as F5-F7. As a result, its power supply capacity becomes constrained in later stages, leading to a growing performance gap compared to ODP.

In summary, although ODP does not show a clear advantage over other methods in the early stages, it accounts for the long-term impact of early-stage repairs on subsequent restoration. This proactive scheduling ensures smoother load restoration progress in later stages, ultimately leading to a significantly higher restoration level around time step 30 compared to other methods.

\begin{figure}[!htbp]
                \centering
                \includegraphics[width=0.8\linewidth]{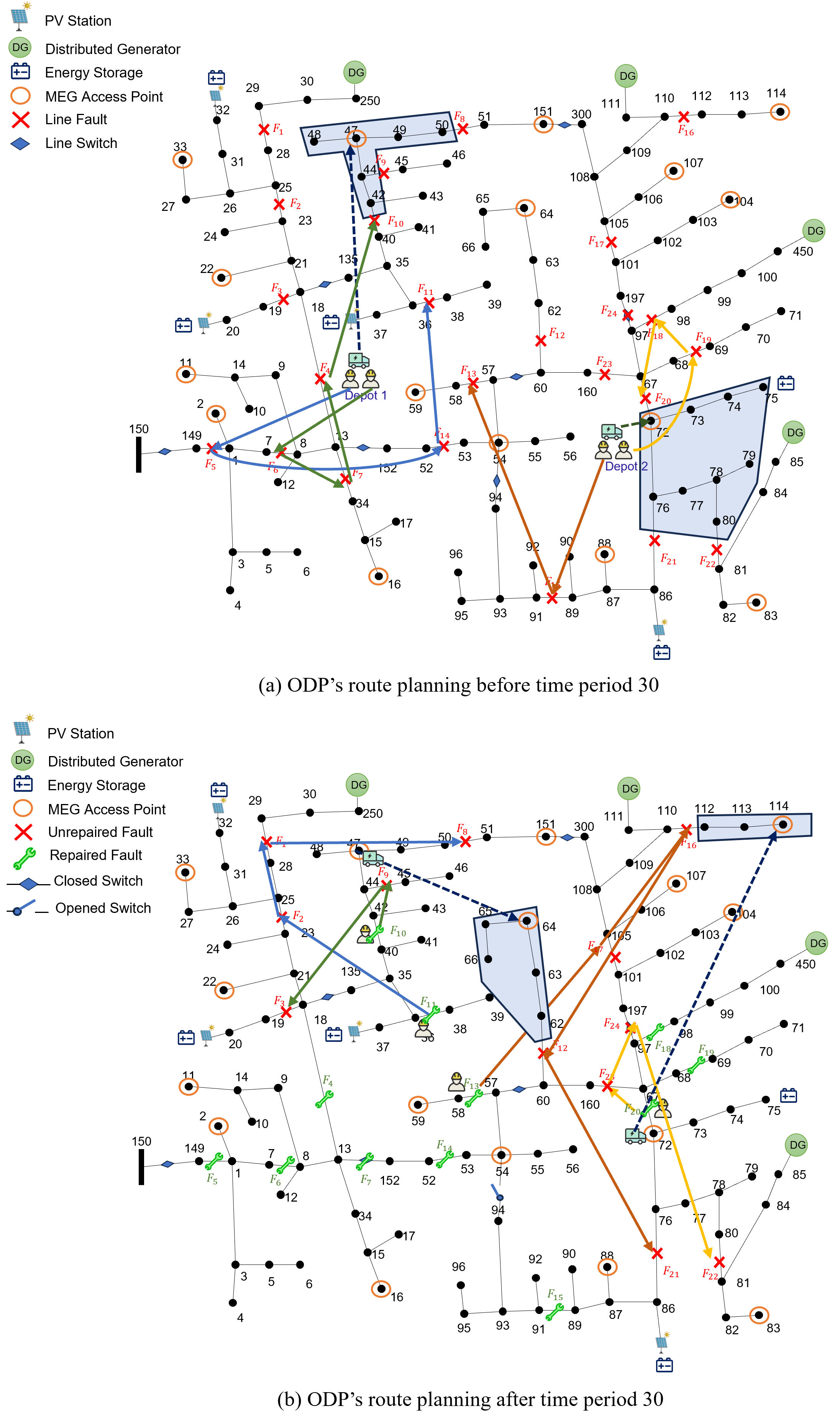}
                \caption{The routing of mobile resources using the proposed ODP method}
                \label{fig:route_123}
        \end{figure}

Fig. \ref{fig:route_123} provides a detailed visualization of the mobile resource routing results of ODP, clearly demonstrating its effectiveness. In the early stages (Fig. \ref{fig:route_123}(a)) of emergency restoration, two repair crews dispatched from Depot 1 collaborated to restore the connection to substation bus 150, ensuring the substation's capacity could be fully utilized to supply downstream buses. Meanwhile, the mobile emergency generator from Depot 1 was deployed to the heavy-load area around bus 47 (Island 10 in Fig. \ref{fig:isl_loss_123}) to supply critical loads. Similarly, the repair crews and the mobile emergency generator from Depot 2 prioritized restoring critical faulty lines and unserved loads near the depot, enabling power from the substation to be further transmitted to the right-side region. This approach balanced repair time with the effectiveness of faulty line restoration.

In the later stages of the restoration process (Fig. \ref{fig:route_123}(b)), the repair crews focused on addressing the remaining faults. By this time, many isolated loads were already being supplied by local distributed generation or mobile emergency generators, reducing the urgency of these faults. This indicates that the ODP method prioritized and restored the more critical lines during the early stages. These results highlight that ODP effectively coordinates various mobile resources across both spatial and temporal dimensions, enabling rapid and efficient emergency restoration. This demonstrates the strong practical value of ODP in post-disaster scenarios.

\begin{figure}[!htbp]
                \centering
                \includegraphics[width=0.68\linewidth]{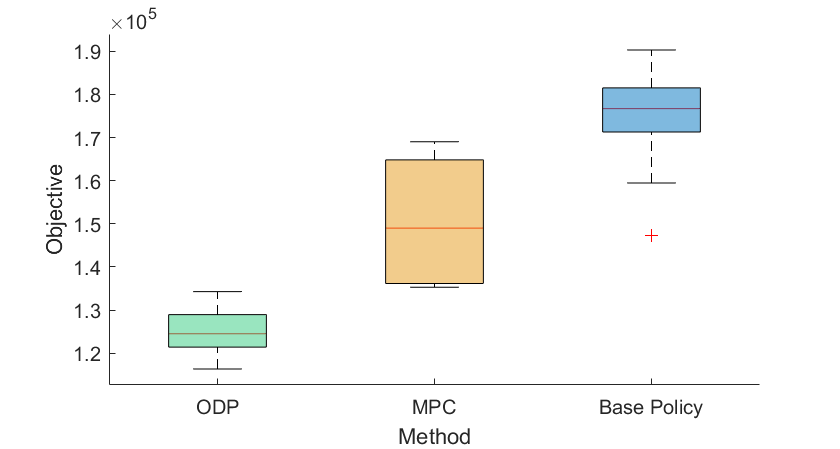}
                \caption{The distribution of cumulative load loss under different experiments}
                \label{fig:boxplot_123}
        \end{figure}

        To further validate the robustness of the ODP strategy, we conducted 20 experiments using different random seeds. These seeds generate different sampled scenarios for ODP and varying uncertainty predictions for MPC and the base policy. The distribution of cumulative load loss across the 20 experiments is illustrated in the boxplot in Fig. \ref{fig:boxplot_123}. The mean values for ODP, MPC, and the base policy are 125,372, 150,632, and 174,589, respectively, with corresponding standard deviations of 4,732, 14,792, and 9,978. ODP exhibits significantly lower variance, highlighting its robustness, attributed to the averaging effect across diverse simulation scenarios. In contrast, MPC shows a much higher variance, as its optimization results are highly sensitive to the predictions of uncertain parameters.

        Since the base policy is an important component of ODP, we also test different heuristics as base policies with the same random seeds. The distance-based policy prioritizes the repair of the nearest lines \cite{base_policy_dist}, and the load-based policy prioritizes the repair of lines that bring the maximum load recovery \cite{base_policy_reco}. Results are shown in Table \ref{table:BP_comp}. The results indicate that online simulation enhances the performance of any base policy, with the proposed base policy outperforming the other two. This highlights the importance of considering both repair efficiency and load recovery when dispatching mobile resources.

        \begin{table}[!htbp]
                \centering
                \scriptsize
                \renewcommand{\arraystretch}{1.5}
                \caption{Comparison of results under different base policies}  \label{table:BP_comp}
                \begin{tabular}{c|c|c}
                        \toprule
                        Base policy  & Load Loss (Base Policy)  & Load loss (ODP) \\
                        \midrule
                        Proposed Base Policy & \$176,235 & \$121,601 \\
                        Distance-Based Policy & \$246,451 & \$152,088 \\ 
                        Load-Based Policy & \$200,062 & \$136,455 \\
                        \hline
                \end{tabular}
        \end{table}
        
        To analyze the impact of the four types of uncertainties, comparative experiments are conducted.  For the proposed ODP method, when only renewable energy and load uncertainties are considered—assuming accurate repair time and fault information—the objective value is \$110,763. This is close to the objective value of \$106,233 under perfect information, suggesting that uncertainties in load demand and renewable energy output have a limited impact. In contrast, when uncertainties in repair time and fault detection are taken into account, while assuming precise values for load and renewable energy outputs, the objective value increases to \$119,792. This finding indicates that uncertainties in repair time and fault detection have a greater impact on results than uncertainties in load demand and renewable energy generation.

        \subsection{Sensitivity Analysis}

        To effectively demonstrate the applicability of the proposed method, it is essential to evaluate its performance under various parameter settings. This section presents the sensitivity analysis from multiple perspectives.

        \subsubsection{Sensitivity Analysis of the Scenario Number}
        
        In the proposed ODP algorithm, $N_s$ scenarios are generated for online parallel simulation, allowing the approximation of Q-values through scenario averaging. The choice of $N_s$ impacts the performance and efficiency of the algorithm: if $N_s$ is too small, the limited number of scenarios may lack representativeness, affecting solution quality; conversely, if $N_s$ is too large, computational time increases.

\begin{figure}[h]
        \centering
        \includegraphics[width=0.6\linewidth]{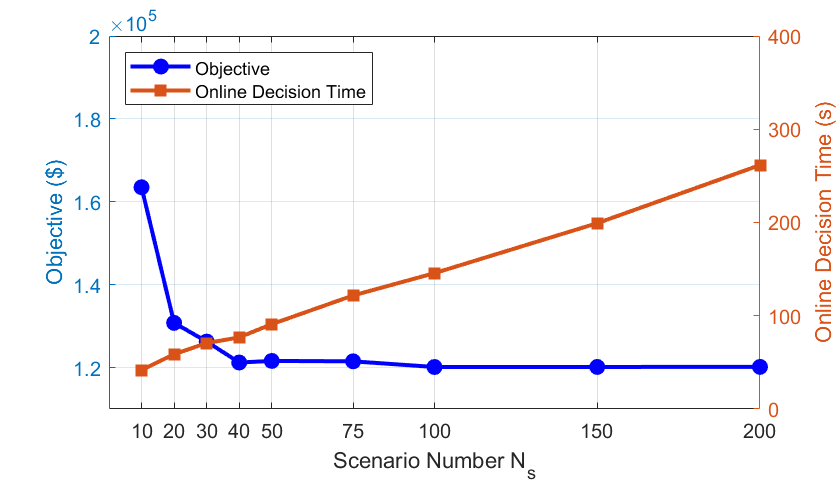}
        \caption{Sensitivity analysis of scenario number}
        \label{fig:sa_ns}
        \end{figure}

        In the 123-bus case study, the baseline experiment uses 50 scenarios. In the revised manuscript, a sensitivity analysis is conducted to examine the impact of different scenario counts, with the results presented in Fig. \ref{fig:sa_ns}. 

        As shown in Fig. \ref{fig:sa_ns}, when the scenario number is below 40, the objective value exhibits fluctuations with increasing scenarios, indicating that 10 or 20 scenarios may not provide sufficient representativeness. However, when the scenario number exceeds 40, the objective value stabilizes, indicating that these scenarios are adequate for approximating expectations. Meanwhile, the online computation time increases with the number of scenarios. Nevertheless, even with 200 scenarios, each online decision update takes less than five minutes, demonstrating the efficiency of the proposed method.

        \subsubsection{Sensitivity analysis of the Action Set Size}

        According to Section 3.3 of the paper, to reduce the computational burden of online simulation, the Q-values \( Q^\pi(s_t, a_t) \) are evaluated on a small subset \( a_t \in \tilde{A_t} \), where \( \tilde{A_t} \) consists of \( K_a \) promising actions selected using the proposed index. However, if \( K_a \) is too small, some truly optimal actions may be excluded, potentially affecting decision quality.  

\begin{figure}[h]
        \centering
        \includegraphics[width=0.6\linewidth]{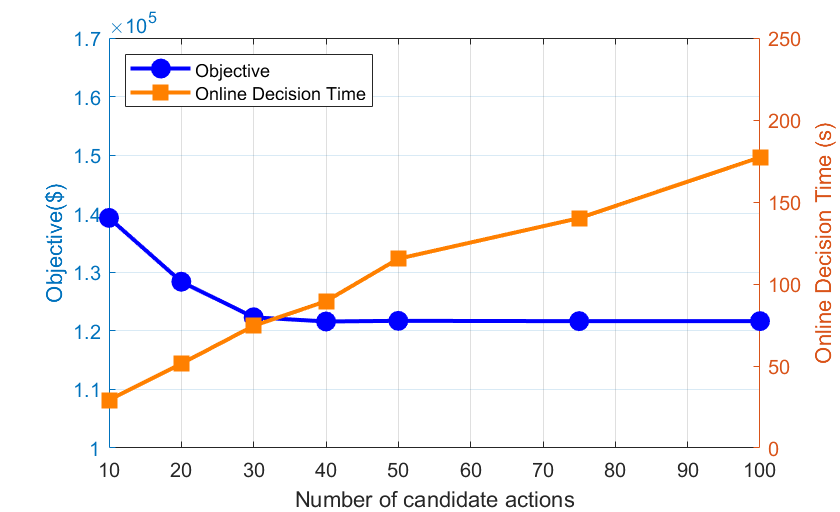}
        \caption{Sensitivity analysis of action set size}
        \label{fig:sa_ka}
        \end{figure}

        In the baseline experiment, the action set size \( K_a \) is set to 40. Fig.~\ref{fig:sa_ka} presents the results for different values of \( K_a \), with the number of online simulation scenarios fixed at 50. The results indicate that the objective stabilizes once \( K_a \) exceeds 30, indicating that this size is sufficient to identify the optimal decision. Moreover, while increasing \( K_a \) leads to longer online decision times, the process remains efficient, with the decision time staying under 5 minutes even when considering 100 candidate actions. This demonstrates the practicality and efficiency of the proposed method.

        \subsubsection{Sensitivity Analysis of the Uncertain Parameter Distribution}

        Since the accuracy of uncertainty distribution influences the online simulation process of the ODP algorithm, a sensitivity analysis is conducted to evaluate model performance under different levels of distribution variance. Table \ref{table:variance} outlines the experimental settings, where the variance of uncertainty distribution gradually increases from Distribution 1 to Distribution 3. For example, the sampling interval for repair time expands from 2 time steps in Distribution 1 to 10 time steps in Distribution 3. Likewise, forecasting errors for renewable energy output and load demand increase from 10\% to 40\%, reflecting the varying degrees of uncertainty in the experiments.

        \begin{table}[h]
        \centering
        \scriptsize
        \renewcommand{\arraystretch}{1.3}
        \caption{Experimental settings with different levels of distribution variance}  \label{table:variance}     
        \begin{tabular}{c|cc}
                \toprule
                &  \makecell{Sampling range for \\ uncertain repair time}  &  \makecell{Forecast error in \\ PV output \& load demand} \\
                \midrule
                Distribution 1 & 2 time steps & 10\%  \\
                Distribution 2 & 6 time steps & 20\% \\
                Distribution 3 & 10 time steps  & 40\% \\
                \hline
        \end{tabular}
        
        \end{table}

\begin{figure}[h]
        \centering
        \includegraphics[width=0.6\linewidth]{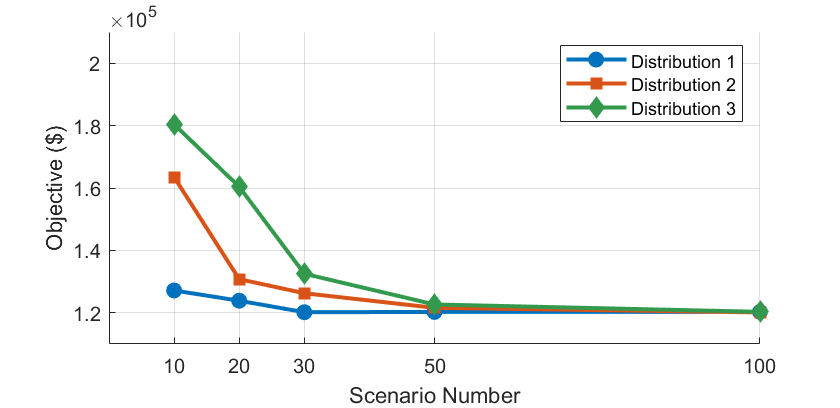}
        \caption{Sensitivity analysis of uncertain parameter distribution}
        \label{fig:sa_distr_2}
        \end{figure}

        The corresponding experimental results are illustrated in Fig. \ref{fig:sa_distr_2}, which demonstrate that a more accurate uncertainty distribution leads to better model performance, even with a limited number of scenarios (e.g., 20 or 30). This indicates that when the uncertainty distribution is precise, even a small number of scenarios can effectively approximate the expected value function. In practical applications, lower uncertainty facilitates more informed decision-making. Moreover, when the number of scenarios exceeds 50, the impact of uncertainty on the results becomes negligible, highlighting the robustness of the proposed method. Given its high computational efficiency, the method can simulate a large number of scenarios simultaneously, leveraging sample average approximation to mitigate the impact of uncertainty.

        \subsection{8500 Bus System}
\begin{figure}[!h]
                \centering
                \includegraphics[width=0.9\linewidth]{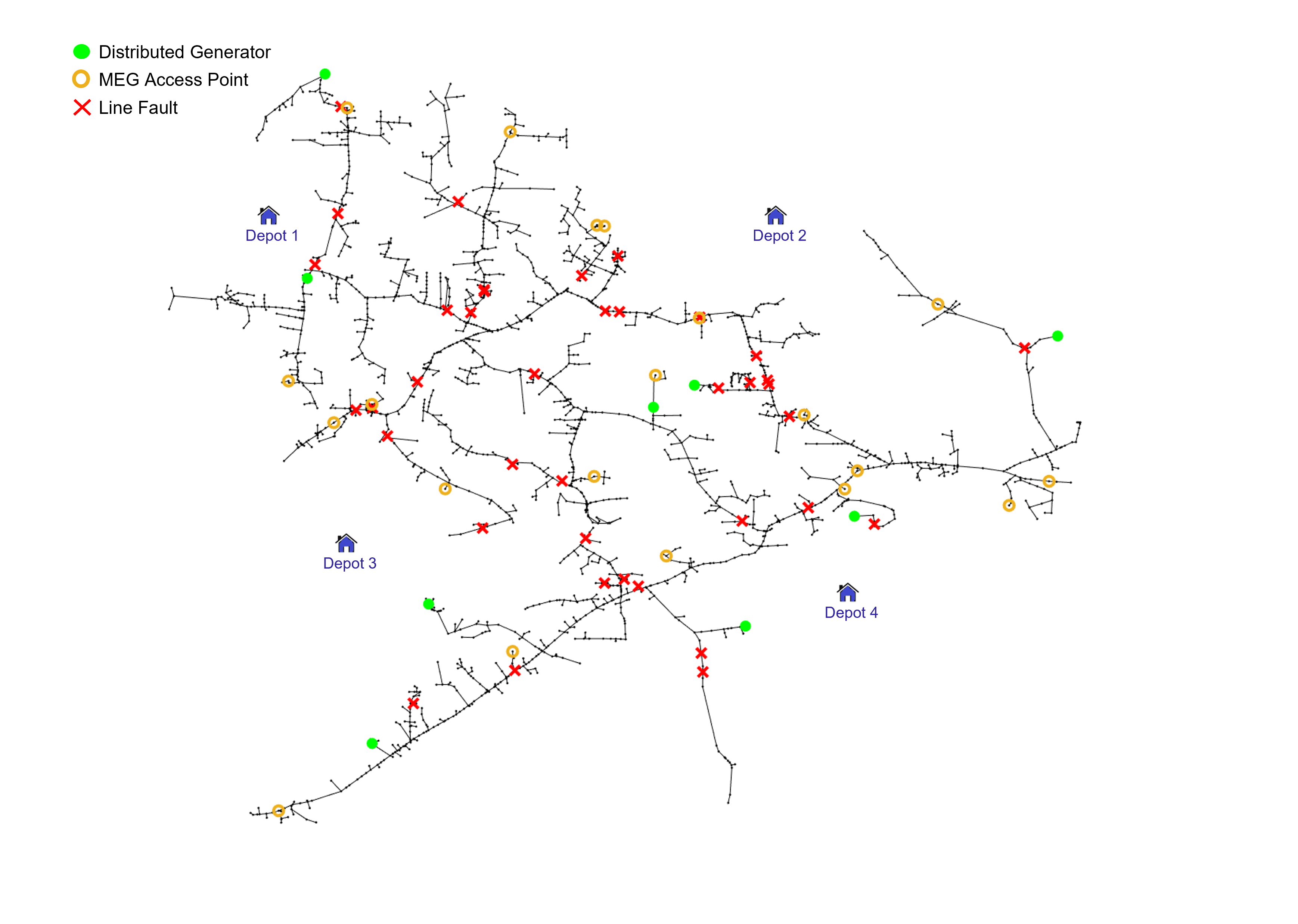}
                \caption{8500-bus system}
                \label{fig:case8500}
                \vspace{0pt}
        \end{figure}

        The testing system with 8500 buses and 40 line faults is shown in Fig. \ref{fig:case8500}. The system includes ten identical micro gas turbines with a total capacity of 8MW, eight repair teams and four mobile emergency generators housed in four depots. The capacity of each mobile generator is 400kW. Due to the larger scale of this system, traveling between fault points takes longer time, so MPC solves a $24$-step lookahead problem which ensures sufficient time to reach distant targets. Due to the scale of the $24$-step lookahead problem, GUROBI is not able to provide a viable solution within $\Delta t = 15$ minutes. As a remedy, the MPC strategy is computed hourly, while the state is updated every 15 minutes. Additionally, due to the extensive integer variables in the all-period optimization problem for the 8500-bus system, the two-stage SP method fails to produce meaningful results and is thus not applied.
        
\begin{figure}[!h]
                \centering
                \includegraphics[width=0.6\linewidth]{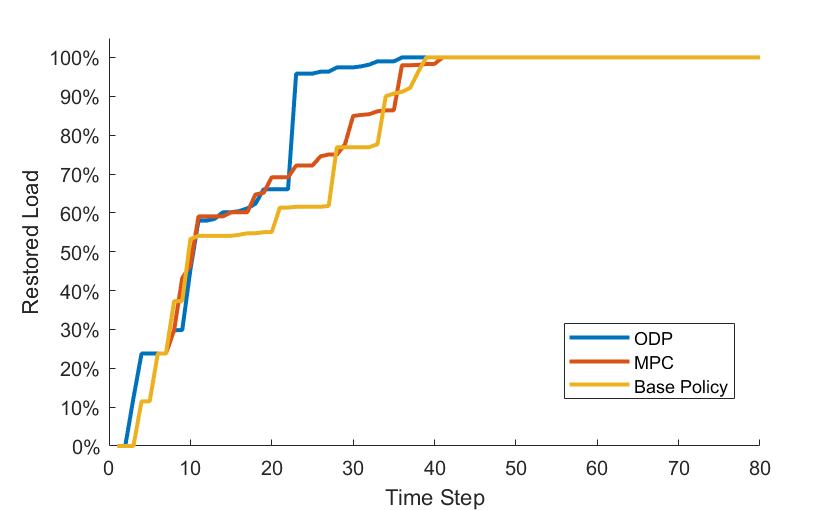}
                \caption{Load restoration for 8500-bus case}
                \label{fig:WLL_8500}
        \end{figure}

        The results of different methods are compared in Table \ref{table:8500_result_comp} and Fig. \ref{fig:WLL_8500}, demonstrating the superiority of ODP. Applying the base policy, a strategy is found in less than 1 second, yielding a cumulative loss of \$277,271. For the $24$-step lookahead problem in MPC, the best feasible solution obtained within one hour results in a 14.7\% improvement over the base policy. In comparison, the proposed ODP method achieves a 24.8\% improvement through comprehensive simulation while maintaining a computation time of less than 5 minutes. This demonstrates the effectiveness and efficiency of ODP in handling large-scale instances. Additionally, since simulations across different scenarios can be executed in parallel, computational efficiency is unlikely to pose a significant barrier in practical applications.
        
        As shown in Fig. \ref{fig:WLL_8500}, ODP achieves faster restoration after period 24 compared to the other methods. As ODP computes Q-value by simulation, its strategy is more foresighted and superior. In contrast, the base policy inherently lacks foresight, while MPC is constrained by computational efficiency and can only achieve a suboptimal solution for a limited horizon.
        
        \begin{table}[!h]
                \centering
                \scriptsize
                \renewcommand{\arraystretch}{1.3}
                \caption{Results of each method on the 8500-bus case}  \label{table:8500_result_comp}     
                \begin{tabular}{c|ccc}
                        \toprule
                        & Objective value  &  Online computation time  & Performance improvement  \\
                        \midrule
                        Base Policy &  \$277,271  & 0.59s & -  \\
                        ODP  & \$208,522  &  289s & 24.8\% \\
                        MPC & \$249,262  & 2041s & 14.7\% \\     
                        \hline
                \end{tabular}
        \end{table}
        
\begin{figure}[!h]
                \centering
                \includegraphics[width=0.6\linewidth]{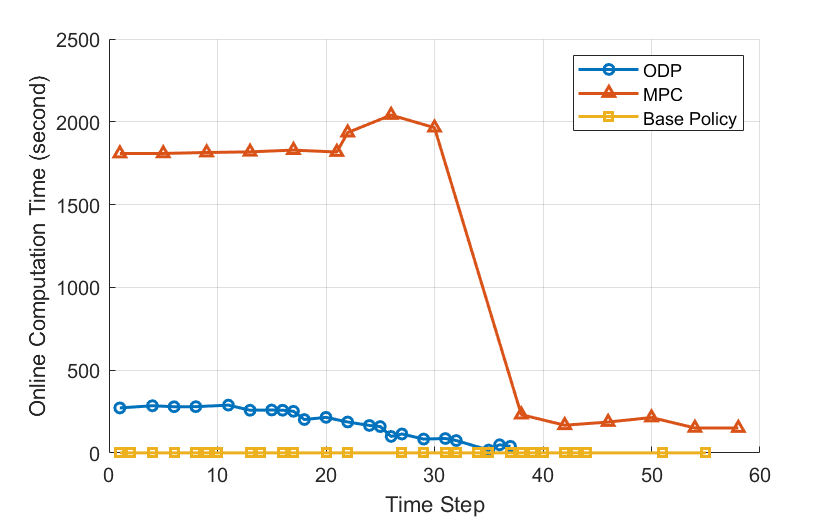}
                \caption{Online computation time for decision updates in the 8500-bus case}
                \label{fig:time_8500}
        \end{figure}

        Fig. \ref{fig:time_8500} shows the online decision-making time for different methods in the 8500-bus system. Each point represents the computation time required for a decision update at a given time step. For ODP and the base policy, updates occur only when the system state or real-time information changes (e.g., when a repair crew completes or starts a task and requires a new target, or when new faults are detected). 
        As a result, online computation time for ODP and the base policy is recorded only when updates are needed, while MPC updates its strategy hourly as previously described.

        As shown in Fig. \ref{fig:time_8500}, ODP consistently completes decision updates within 5 minutes, with computation time decreasing in later stages as fewer faults remain. The base policy updates decisions in under 1 second. In contrast, MPC is significantly less efficient, requiring over 30 minutes for updates in the early stages. This inefficiency stems from its use of a brute-force branch-and-bound algorithm to solve the mixed-integer programming problem, which is highly computationally intensive for large-scale mobile resource routing. Consequently, MPC is not suitable for real-time decision-making in large-scale scenarios.

        \section{Conclusion} \label{section: conclusion}
        
        This paper presents a simulation-based decision-making method for distribution system restoration utilizing mobile resources. The proposed method eliminates the need for offline model training or solving large-scale optimization problems, making it well-suited for real-time online decision-making. The main conclusions are as follows:

        -- High Computational Efficiency and Online Adaptability:  The proposed method enables real-time decision-making within 5 minutes for both the 123-bus and 8500-bus cases, demonstrating strong online adaptability in mobile resource scheduling.

        -- Effective Restoration Performance: Compared to the base policy, the ODP policy enhances load restoration by over 20\%, highlighting its effectiveness in post-disaster restoration.

        -- Robustness Against Uncertainty: Due to its high computational efficiency, the ODP method can evaluate promising actions under a large number of sampled scenarios, ensuring reasonable action selection even with imperfect uncertainty information.

        Future research could focus on integrating the proposed ODP method with learning-based approaches to further enhance decision-making when trained models are available. Additionally, extending its application to integrated energy system restoration could be another promising direction.

        \bibliographystyle{elsarticle-num} 
        \bibliography{refs_3}

@misc{MEG_hurricane_sandy,
  author = {{FEMA}},
  title = {Mitigation Assessment Team Report - Hurricane Sandy in New Jersey and New York},
  year = {2013},
  howpublished = {Available: \url{https://www.govinfo.gov/content/pkg/GOVPUB-HS5-PURL-gpo110213/pdf/GOVPUB-HS5-PURL-gpo110213.pdf}}
}

@article{pre_disaster_SP_1,
  title={Mobile emergency generator pre-positioning and real-time allocation for resilient response to natural disasters},
  author={Lei, Shunbo and Wang, Jianhui and Chen, Chen and Hou, Yunhe},
  journal={IEEE Trans. Smart Grid},
  volume={9},
  number={3},
  pages={2030--2041},
  year={2016},
  publisher={IEEE}
}

@article{pre_disaster_SP_3,
  title={Preventive allocation and post-disaster cooperative dispatch of emergency mobile resources for improved distribution system resilience},
  author={Shi, Qingxin and Wan, Haiyang and Liu, Wenxia and Han, Haiteng and Wang, Zhuorong and Li, Fangxing},
  journal={Int J Electr Power Energy Syst},
  volume={152},
  pages={109238},
  year={2023},
  publisher={Elsevier}
}

@article{pre_disaster_RO_2,
  title={A distributionally robust resilience enhancement model for transmission and distribution coordinated system using mobile energy storage and unmanned aerial vehicle},
  author={Wang, Zekai and Ding, Tao and Mu, Chenggang and Huang, Yuhan and Yang, Miao and Yang, Yueyang and Lin, Yi and Li, Meng},
  journal={Int J Electr Power Energy Syst},
  volume={152},
  pages={109256},
  year={2023},
  publisher={Elsevier}
}

@article{RO_1,
  title={Robust coordination of repair and dispatch resources for post-disaster service restoration of the distribution system},
  author={Wu, Hao and Xie, Yunyun and Xu, Yan and Wu, Qiuwei and Yu, Chen and Sun, Jinsheng},
  journal={Int. J. Elect. Power Energy Syst.},
  volume={136},
  pages={107611},
  year={2022},
  publisher={Elsevier}
}

@article{RO_SP_1,
  title={Hybrid modeling based co-optimization of crew dispatch and distribution system restoration considering multiple uncertainties},
  author={Li, Jiayong and Khodayar, Mohammad E and Feizi, Mohammad Ramin},
  journal={IEEE Syst. J.},
  volume={16},
  number={1},
  pages={1278--1288},
  year={2021},
  publisher={IEEE}
}

@article{Decomposed_SP_1,
  title={Optimizing service restoration in distribution systems with uncertain repair time and demand},
  author={Arif, Anmar and Ma, Shanshan and Wang, Zhaoyu and Wang, Jianhui and Ryan, Sarah M and Chen, Chen},
  journal={IEEE Trans. Power Syst.},
  volume={33},
  number={6},
  pages={6828--6838},
  year={2018},
  publisher={IEEE}
}

@article{nonantcp,
  title={Multistage transmission-constrained unit commitment with renewable energy and energy storage: implicit and explicit decision methods},
  author={Zhou, Yuzhou and Zhai, Qiaozhu and Wu, Lei},
  journal={IEEE Trans. Sustain. Energy},
  volume={12},
  number={2},
  pages={1032--1043},
  year={2020},
  publisher={IEEE}
}

@article{MPC_SP_1,
  title={Rolling optimization of mobile energy storage fleets for resilient service restoration},
  author={Yao, Shuhan and Wang, Peng and Liu, Xiaochuan and Zhang, Huajun and Zhao, Tianyang},
  journal={IEEE Trans. Smart Grid},
  volume={11},
  number={2},
  pages={1030--1043},
  year={2019},
  publisher={IEEE}
}

@article{MPC_2,
  title={Collaborative distribution system restoration planning and real-time dispatch considering behind-the-meter DERs},
  author={Liu, Weijia and Ding, Fei},
  journal={IEEE Trans. Power Syst.},
  volume={36},
  number={4},
  pages={3629--3644},
  year={2020},
  publisher={IEEE}
}

@article{DRO_1,
  title={Decentralized data-driven load restoration in coupled transmission and distribution system with wind power},
  author={Zhao, Jin and Wu, Qiuwei and Hatziargyriou, Nikos D and Li, Fangxing and Teng, Fei},
  journal={IEEE Trans. Power Syst.},
  volume={36},
  number={5},
  pages={4435--4444},
  year={2021},
  publisher={IEEE}
}

@ARTICLE{DRO_2,
  author={Lu, Shuai and Li, Yuan and Ding, Shixing and Gu, Wei and Xu, Yijun and Song, Meng},
  journal={IEEE Trans. Ind. Inform.}, 
  title={Combined Electrical and Heat Load Restoration Based on Bi-Objective Distributionally Robust Optimization}, 
  year={2023},
  volume={19},
  number={8},
  pages={9239-9252},
  publisher={IEEE}
}

@article{SP_PH_1,
  title={Two-stage service restoration of integrated electric and heating system with the support of mobile heat sources},
  author={Shi, Han and Xie, Yunyun and Hou, Kai and Cai, Sheng and Jia, Hongjie and Wu, Hao and Sun, Jinsheng},
  journal={Appl. Energy},
  volume={379},
  pages={124899},
  year={2025},
  publisher={Elsevier}
}

@article{RL_1,
  title={Curriculum-based reinforcement learning for distribution system critical load restoration},
  author={Zhang, Xiangyu and Eseye, Abinet Tesfaye and Knueven, Bernard and Liu, Weijia and Reynolds, Matthew and Jones, Wesley},
  journal={IEEE Trans. Power Syst.},
  year={2022},
  publisher={IEEE}
}

@article{RL_2,
  title={Hybrid imitation learning for real-time service restoration in resilient distribution systems},
  author={Zhang, Yichen and Qiu, Feng and Hong, Tianqi and Wang, Zhaoyu and Li, Fangxing},
  journal={IEEE Trans. Ind. Inform.},
  volume={18},
  number={3},
  pages={2089--2099},
  year={2021},
  publisher={IEEE}
}

@article{RL_RC,
  title={Hierarchical multi-agent reinforcement learning for repair crews dispatch control towards multi-energy microgrid resilience},
  author={Qiu, Dawei and Wang, Yi and Zhang, Tingqi and Sun, Mingyang and Strbac, Goran},
  journal={Appl. Energy},
  volume={336},
  pages={120826},
  year={2023},
  publisher={Elsevier}
}

@article{DL_MCTS_RC,
  title={Post-storm repair crew dispatch for distribution grid restoration using stochastic Monte Carlo tree search and deep neural networks},
  author={Shuai, Hang and Li, Fangxing and She, Buxin and Wang, Xiaofei and Zhao, Jin},
  journal={Int J Electr Power Energy Syst},
  volume={144},
  pages={108477},
  year={2023},
  publisher={Elsevier}
}

@article{RL_MESS,
  title={Multi-agent deep reinforcement learning for resilience-driven routing and scheduling of mobile energy storage systems},
  author={Wang, Yi and Qiu, Dawei and Strbac, Goran},
  journal={Appl. Energy},
  volume={310},
  pages={118575},
  year={2022},
  publisher={Elsevier}
}

@article{RL_4,
  title={Deep reinforcement learning-based model-free on-line dynamic multi-microgrid formation to enhance resilience},
  author={Zhao, Jin and Li, Fangxing and Mukherjee, Srijib and Sticht, Christopher},
  journal={IEEE Trans. Smart Grid},
  volume={13},
  number={4},
  pages={2557--2567},
  year={2022},
  publisher={IEEE}
}

@article{RL_topo_2,
  title={A multi-agent reinforcement learning method for distribution system restoration considering dynamic network reconfiguration},
  author={Si, Ruiqi and Chen, Siyuan and Zhang, Jun and Xu, Jian and Zhang, Luxi},
  journal={Appl. Energy},
  volume={372},
  pages={123625},
  year={2024},
  publisher={Elsevier}
}

@article{RL_6,
  title={Distribution system resilience under asynchronous information using deep reinforcement learning},
  author={Bedoya, Juan Carlos and Wang, Yubo and Liu, Chen-Ching},
  journal={IEEE Trans. Power Syst.},
  volume={36},
  number={5},
  pages={4235--4245},
  year={2021},
  publisher={IEEE}
}

@article{RL_7,
  title={Learning sequential distribution system restoration via graph-reinforcement learning},
  author={Zhao, Tianqiao and Wang, Jianhui},
  journal={IEEE Trans. Power Syst.},
  volume={37},
  number={2},
  pages={1601--1611},
  year={2021},
  publisher={IEEE}
}

@book{bertsekas_book,
  title={Rollout, policy iteration, and distributed reinforcement learning},
  author={Bertsekas, Dimitri},
  year={2021},
  publisher={Athena Scientific}
}

@article{uncertain_repair_time_2,
  title={Restoration of high-renewable-penetrated distribution systems considering uncertain repair workloads},
  author={Sun, Xiaotian and Xie, Haipeng and Bie, Zhaohong and Li, Gengfeng},
  journal={CSEE J. Power Energy Syst.},
  year={2022},
  publisher={CSEE},
  doi={10.17775/CSEEJPES.2021.06500}
}

@article{base_policy_reco,
  title={Dynamic repair scheduling for transmission systems based on look-ahead strategy approximation},
  author={Yan, Jiahao and Hu, Bo and Xie, Kaigui and Niu, Tao and Li, Chunyan and Tai, Heng-Ming},
  journal={IEEE Trans. Power Syst.},
  volume={36},
  number={4},
  pages={2918--2933},
  year={2020},
  publisher={IEEE}
}

@article{base_policy_dist,
  title={Resiliency-based optimization of restoration policies for electric power distribution systems},
  author={Figueroa-Candia, Marcelo and Felder, Frank A and Coit, David W},
  journal={Electr. Power Syst. Res.},
  volume={161},
  pages={188--198},
  year={2018},
  publisher={Elsevier}
}

@article{resilience_review_1,
  title={A systematic review on power system resilience from the perspective of generation, network, and load},
  author={Wang, Chong and Ju, Ping and Wu, Feng and Pan, Xueping and Wang, Zhaoyu},
  journal={Renew. Sustain. Energy Rev.},
  volume={167},
  pages={112567},
  year={2022},
  publisher={Elsevier}
}

@article{MES_6,
  title={Two-stage mobile emergency generator dispatch for sequential service restoration of microgrids in extreme conditions},
  author={Cai, Sheng and Xie, Yunyun and Wu, Qiuwei and Jin, Xiaolong and Zhang, Menglin and Xiang, Zhengrong},
  journal={Int J Electr Power Energy Syst},
  volume={153},
  pages={109312},
  year={2023},
  publisher={Elsevier}
}

@article{MIP_road_2,
  title={Co-optimize recovery modeling for transportation and power network with multi-type mobile resources dispatching},
  author={Sun, Shaohua and Li, Gengfeng and Yang, Qiming and Bie, Zhaohong},
  journal={Applied Energy},
  volume={366},
  pages={123074},
  year={2024},
  publisher={Elsevier}
}

@article{MIP_MEG_1,
  title={Resilient service restoration for unbalanced distribution systems with distributed energy resources by leveraging mobile generators},
  author={Ye, Zhigang and Chen, Chen and Chen, Bo and Wu, Kai},
  journal={IEEE Trans. Ind. Inform.},
  volume={17},
  number={2},
  pages={1386--1396},
  year={2020},
  publisher={IEEE}
}

@article{MIP_EB_1,
  title={Routing and scheduling of electric buses for resilient restoration of distribution system},
  author={Li, Boda and Chen, Ying and Wei, Wei and Huang, Shaowei and Xiong, Yufeng and Mei, Shengwei and Hou, Yunhe},
  journal={IEEE Trans. Transp. Electrif.},
  volume={7},
  number={4},
  pages={2414--2428},
  year={2021},
  publisher={IEEE}
}

@article{MIP_EB_3,
  title={A collaborative restoration strategy of resilient distribution system with the support of electric bus clusters},
  author={Guo, Yina and Liao, Kai and Yang, Jianwei and Zheng, Shunwei and He, Zhengyou},
  journal={Int J Electr Power Energy Syst},
  volume={161},
  pages={110199},
  year={2024},
  publisher={Elsevier}
}

@article{MIP_line_repair_1,
  title={Multiperiod distribution system restoration with routing repair crews, mobile electric vehicles, and soft-open-point networked microgrids},
  author={Ding, Tao and Wang, Zekai and Jia, Wenhao and Chen, Bo and Chen, Chen and Shahidehpour, Mohammad},
  journal={IEEE Trans. Smart Grid},
  volume={11},
  number={6},
  pages={4795--4808},
  year={2020},
  publisher={IEEE}
}

@article{MIP_line_repair_2,
  title={Sequential disaster recovery model for distribution systems with co-optimization of maintenance and restoration crew dispatch},
  author={Zhang, Gang and Zhang, Feng and Zhang, Xin and Meng, Ke and Dong, Zhao Yang},
  journal={IEEE Trans. Smart Grid},
  volume={11},
  number={6},
  pages={4700--4713},
  year={2020},
  publisher={IEEE}
}

@article{radial_SCF,
  title={On the radiality constraints for distribution system restoration and reconfiguration problems},
  author={Wang, Ying and Xu, Yin and Li, Jiaxu and He, Jinghan and Wang, Xiaojun},
  journal={IEEE Trans. Power Syst.},
  volume={35},
  number={4},
  pages={3294--3296},
  year={2020},
  publisher={IEEE}
}

@InProceedings{ODP_robot,
  title = 	 {Multiagent Rollout and Policy Iteration for POMDP with Application to Multi-Robot Repair Problems},
  author =       {Bhattacharya, Sushmita and Kailas, Siva and Badyal, Sahil and Gil, Stephanie and Bertsekas, Dimitri},
  booktitle = 	 {Proceedings of the 2020 Conference on Robot Learning},
  pages = 	 {1814--1828},
  year = 	 {2021},
  editor = 	 {Kober, Jens and Ramos, Fabio and Tomlin, Claire},
  volume = 	 {155},
  series = 	 {Proceedings of Machine Learning Research},
  month = 	 {16--18 Nov},
  publisher =    {PMLR},
}

@book{Rollout_Book,
  title={Rollout, policy iteration, and distributed reinforcement learning},
  author={Bertsekas, Dimitri},
  year={2021},
  publisher={Athena Scientific}
}

@article{DistFlow,
  title={Optimal capacitor placement on radial distribution systems},
  author={Baran, Mesut E and Wu, Felix F},
  journal={IEEE Trans. Power Deliv.},
  volume={4},
  number={1},
  pages={725--734},
  year={1989},
  publisher={IEEE}
}

@article{pdf_repair_time,
  title={Optimum post-disruption restoration under uncertainty for enhancing critical infrastructure resilience},
  author={Fang, Yi-Ping and Sansavini, Giovanni},
  journal={Reliab. Eng. Syst. Saf.},
  volume={185},
  pages={1--11},
  year={2019},
  publisher={Elsevier}
}

@article{pdf_solar,
  title={Enhanced performance Gaussian process regression for probabilistic short-term solar output forecast},
  author={Najibi, Fatemeh and Apostolopoulou, Dimitra and Alonso, Eduardo},
  journal={Int. J. Electr. Power Energy Syst.},
  volume={130},
  pages={106916},
  year={2021},
  publisher={Elsevier}
}

@misc{data_github,
  author = {M. Li},
  title = {Data for distribution network restoration},
  year = 2023,
  url = "https://github.com/kxxs/Data_ODP_Restoration"
}
        
\end{document}